\documentclass[superscriptaddress, groupedaddress, twocolumn, amsmath,amssymb, aps, prl]{revtex4}
\usepackage{graphicx}
\usepackage{bm}
\usepackage{color}
\usepackage[colorlinks=true ,citecolor=blue,urlcolor=blue]{hyperref}

\def\micron{{\ \mu{\rm m}}}					

\newcommand{\ket}[1]{|#1\rangle}

\begin{document}

\title{Vortex nucleations in spinor Bose condensates under localized synthetic magnetic fields}

\author{L.~-R. Liu}\thanks{These authors contributed equally to this work.}
\author{S.~-C. Wu}\thanks{These authors contributed equally to this work.}
\author{T.~-W. Liu}
\author{H.~-Y. Hsu}
\author{T.~-K. Shen}
\affiliation{Institute of Atomic and Molecular Sciences, Academia Sinica, Taipei, Taiwan 10617}
\author{S.~-K. Yip}
\affiliation{Institute of Atomic and Molecular Sciences, Academia Sinica, Taipei, Taiwan 10617}
\affiliation{Institute of Physics,Academia Sinica, Taipei, Taiwan 11529}
\author{Y. Kawaguchi}
\affiliation{Department of Applied Physics, Nagoya University, Nagoya, 464-8603, Japan}
\affiliation{Research Center for Crystalline Materials Engineering, Nagoya University, Nagoya 464-8603, Japan}
\author{Y.~-J. Lin}
\email{linyj@as.edu.tw}
\affiliation{Institute of Atomic and Molecular Sciences, Academia Sinica, Taipei, Taiwan 10617}
\affiliation{Department of Physics, National Tsing Hua University, Hsinchu 30013, Taiwan}

\date{\today}

\begin{abstract}
Gauge fields are ubiquitous in modern quantum physics. In superfluids, quantized vortices can be induced by gauge fields. Here we demonstrate the first experimental observation of vortex nucleations in light-dressed spinor Bose-Einstein condensates under radially-localized synthetic magnetic fields. The light-induced spin-orbital-angular-momentum coupling creates azimuthal gauge potentials $\vec{A}$ for the lowest-energy spinor branch dressed eigenstate. 
The observation of the atomic wave function in the lowest-energy dressed eigenstate reveals that vortices nucleate from the cloud center of a vortex-free state with canonical momentum $\vec{p} = 0$. This is because a large circulating azimuthal velocity field $\propto \vec{p}-\vec{A}$ at the condensate center results in a dynamically unstable localized excitation that initiates vortex nucleations.	Furthermore, the long-time dynamics to reach the ground state stops in a metastable state when $|\vec{A}|$ is not sufficiently large. Our observation has reasonable agreement with the time-dependent Gross-Pitaevskii simulations.

\end{abstract}

\maketitle

The realization of synthetic gauge fields for charge-neutral ultracold atoms has opened new opportunities for creating and investigating topological quantum matters in a clean and easy-to-manipulate environment~\cite{Dalibard11,Galitski2013,Goldman2014,Zhai2015,Madison00,Chevy2000,Aboshaeer01,schweikhard04}. Early pioneering experiments utilize mechanical rotation to create an effective Lorentz force and thus effective magnetic fields in the rotating frame, realizing lattices of more than a hundred vortices. However, mechanical rotations create only a uniform effective magnetic field $B^*$ with a synthetic gauge potential restricted along the azimuthal direction, $\vec{A}=B^*(-y{\mathbf e}_{x}+x {\mathbf e}_{y})/2$. A breakthrough was achieved by the realization of laser-engineered synthetic gauge potentials in the laboratory frame, which enabled us to implement the Landau gauge $\vec{A}= -B^*y {\mathbf e}_{x}$~\cite{Lin09b}, opening the possibility of engineering more flexible forms of $\vec{A}$.

Under such a variety of synthetic gauge fields, a question naturally arises as to how vortices are nucleated. Quantized vortices, signified as a quantized circulation originating from a single-valuedness of the superfluid order parameters, are widely investigated in quantum condensed systems, such as helium superfluids, superconductors, and neutron stars. Vortex nucleations are of particular interest as they manifest transitions between different topological states. In atomic quantum gases, early works investigated vortex nucleations in mechanically stirred rotating scalar Bose-Einstein condensates (BECs), including experiments~\cite{Madison00,Madison01,Chevy2002,Haljan2001,Aboshaeer01,Hodby2001} and simulations~\cite{Sinha2001,Lobo2004,Dalfovo2000,Simula2002,Fetter2009,Kasamatsu2003}. When a BEC with a negligibly small amount of thermal atoms is stirred with a rotating elliptical trapping potential~\cite{Madison00,Madison01,Chevy2002}, vortices are nucleated from the edge of the BEC owing to dynamical instabilities occurring at the rotating frequency nearly resonant with that of the surface quadrupole mode. This critical frequency is significantly higher than that for thermodynamically stable single-vortex~\cite{Lundh1997}. The $2\pi$ phase slips between quantized supercurrents in ring-shaped quantum gases are also studied in Refs.~\cite{Ramanathan2011,Moulder2012,Beattie2013,Cai2022,DelPace2022}.

Light-induced synthetic magnetic fields with the Landau gauge can also nucleate vortices, as studied experimentally~\cite{Lin09b,LeBlanc2015,Price2016} and theoretically~\cite{Taylor2011}. Here the synthetic field arises from the coupling between the atoms' internal spin and the center-of-mass linear momentum provided by Raman laser dressing, a spin-orbit coupling. In the experiments~\cite{Lin09b,LeBlanc2015}, vortices are observed to appear from the 
edge when the initial vortex-free system is thermodynamically unstable. More recently, physicists also realize synthetic magnetic fields under azimuthal gauge potentials~\cite{Chen2018,Chen2018a,Zhang2019}. This is achieved by coupling the atomic internal spin states and the center-of-mass 
orbital-angular-momentum (OAM), which we refer to as spin-OAM coupling (SOAMC).

In this Letter, we report the first experimental observation of vortex nucleation in a spinor BEC under SOAMC, which creates a vector potential $\vec{A}$ for the lowest-energy Raman dressed state. The condensate wave function of this state is $\varphi$. The synthetic magnetic field $\vec{B}=\vec{\nabla} \times \vec{A}$ is localized around $r\sim 0$ (Fig.~1c) in an almost cylindrically-symmetric system with the coordinate $(r,\phi,z)$. It creates a nonzero circulating kinetic velocity field $(\hbar\vec{\nabla} \vartheta-\vec{A})/m$ even in a vortex-free system with $\vec{\nabla}\vartheta\approx 0$, where $m$ is the atomic mass and $\vartheta$ is the phase of $\varphi$. In the experiment, we adiabatically turn on the gauge field, hold the system for some time, and then adiabatically turn off the gauge field to probe the change in the profile of $\vartheta$ via density images and OAM measurements. The azimuthal velocity under $\vec{A}$ has a maximal value at small $r$ (Fig.~1b) compared to the BEC's radius, and when it exceeds the critical velocity, a mode localized at $r\sim 0$ has a negative energy. When this mode couples with another positive-energy excitation, dynamical instability arises and trigger vortex nucleation. The signature of this unstable mode is observed as vortex-antivortex-pair generation near the center in $\varphi$, see Fig.~5c2. The vortex nucleation proceeds with asymmetry and dissipation, essential for violating OAM and energy conservations. Our vortex nucleation under a spatially localized $\vec{B}$ has drastically different features in the initial instability from those with uniform and radially increasing $\vec{B}$~\cite{Murray2007,Murray2009}, where vortices enter from the edge due to unstable surface modes.

\begin{figure}
	\centering
	\includegraphics[width=3.45 in]{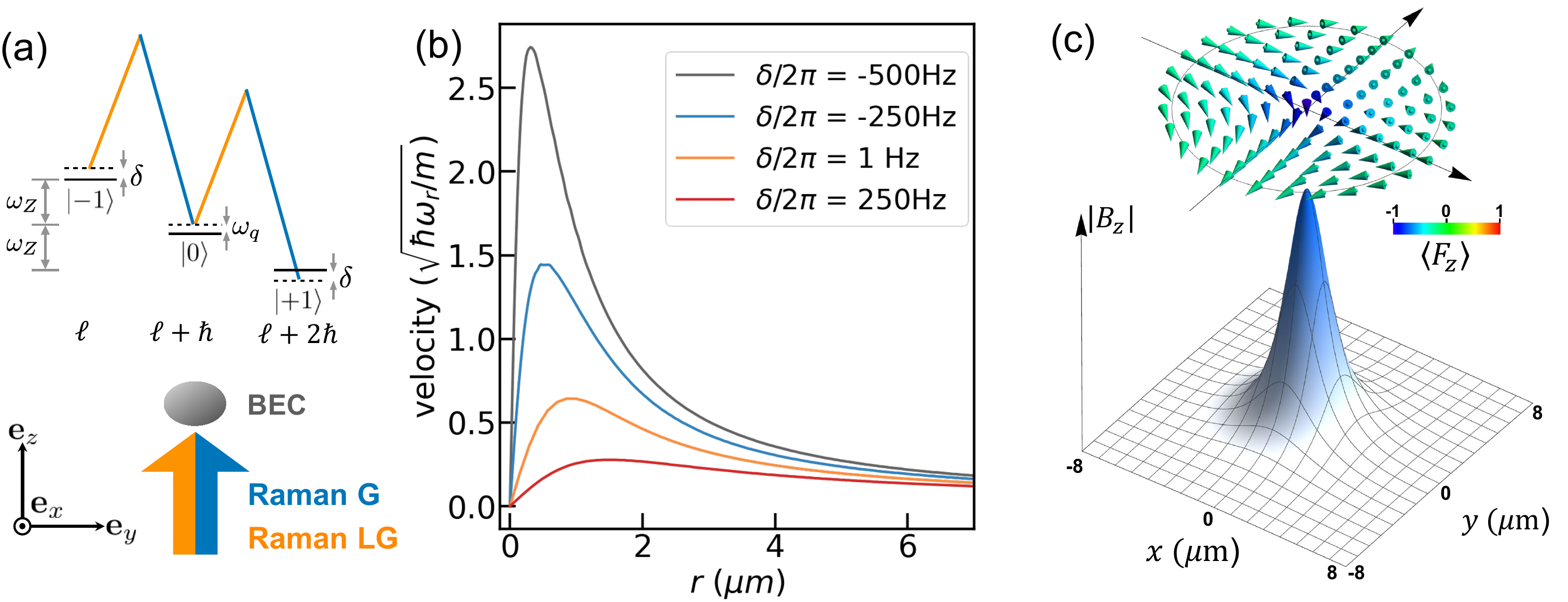}
	\caption{(a) Schematic of SOAMC. (b) Azimuthal velocity vs. $r$ of the Gross-Pitaevskii ground state with $\ell=0$ for various detunings $\delta$. (c) spin texture $\langle \vec{F} \rangle$ in the lowest-energy Raman-dressed state where the color scale indicates $\langle F_z \rangle$, and resulting localized synthetic magnetic field $|\vec{B}|$ where $\vec{B}=\nabla \times A_{-1}\hat{\phi}=B_z\hat{z}$ at $\delta/2\pi=500$~Hz.}
\end{figure}

We implement the gauge potential by loading a spin $F=1$ $^{87}$Rb BEC into the lowest-energy branch of the Raman-dressed states~\cite{Chen2018a}, where a Gaussian Raman beam and a Laguerre-Gaussian (LG) Raman beam with phase winding $\Delta \ell/\hbar=1$ transfer OAM of $\pm \hbar$ when coupling the bare spin state $|m_F\rangle$ to $|m_F\pm 1\rangle$ (Fig.~1a). These Raman beams introduce the SOAMC, $H_\Omega=\vec{\Omega}_\textrm{eff}(\vec{r})\cdot \vec{F}$, where $\vec{F}$ is the vector of spin operators and $\vec{\Omega}_\textrm{eff}=\Omega(r)\cos\phi {\bf e}_x-\Omega(r)\sin\phi {\bf e}_y+\delta {\bf e}_z$ with  $\delta$ being the Raman detuning and $\Omega(r)=\Omega_M\sqrt{e}(r/r_M)e^{-r^2/2r^2_M}$ the Raman coupling strength. Our experiment has $\Omega_M/2\pi=2.5(2)$~kHz and $r_M=17~\mu$m.
The quadratic Zeeman shift is $\omega_q/2\pi\approx 50$~Hz. Because $H_\Omega$ dominates the spin-dependent part of the Hamiltonian, it is convenient to introduce the dressed spin basis $|\xi_{n=0,\pm1}(\vec{r})\rangle$ defined by $H_\Omega|\xi_n(\vec{r})\rangle=n|\vec{\Omega}_\textrm{eff}(\vec{r})||\xi_n(\vec{r})\rangle$; During the vortex nucleation dynamics, the spinor order parameter is well approximated as $\varphi(\vec{r})|\xi_{-1}(\vec{r})\rangle$ with a scalar wave function $\varphi(\vec{r})$. Here, $|\xi_n\rangle$ has a phase ambiguity, and we define $|\xi_{-1}\rangle=\left(e^{i2\phi}\frac{1-\cos\beta}{2}, -e^{i\phi}\frac{\sin\beta}{\sqrt{2}}, \frac{1+\cos\beta}{2}\right)^\textrm{T}$ with $\tan\beta=\Omega(r)/\delta$ so that the initial $\varphi$ is vortex-free, where $\textrm{T}$ denotes transpose. The spatial dependence of $|\xi_{-1}\rangle$ creates the gauge potential $\vec{A}=i\hbar\langle \xi_{-1}|\vec{\nabla}|\xi_{-1}\rangle=A_{-1}(r)\hat{\phi}$ with $rA_{-1}=\hbar(\cos\beta-1)$ effectively acting on $\varphi$. Our interest is the dynamics of $\varphi$ under this $\vec{A}$.

Using $\vec{A}$, the mechanical OAM of the system, which is the population-weighted summation of the OAM of each spin $m_F$ component, is given by $\int d\vec{r}\varphi^*(\hat{\ell}-rA_{-1})\varphi$ where $\hat{\ell}=-i\hbar\partial_\phi$. Thus, even when we start from a vortex-free $\varphi$, the system would evolve to a state whose canonical OAM (cOAM) $\tilde{L}_z=\int d\vec{r}\varphi^*\hat{\ell}\varphi$ becomes nonzero. Indeed, $\varphi$ in the ground state is the eigenstate of $\hat{\ell}$ with the eigenvalue $\ell_g=0,-\hbar,-2\hbar$ for $\delta/2\pi>200$~Hz, $|\delta/2\pi|<200$~Hz, and $\delta/2\pi<-200$~Hz, respectively~\cite{Chen2018a}, so the initial condensate at $\delta/2\pi<200$~Hz with $\ell=0\neq \ell_g$ are expected to temporally evolve if there are instabilities. Note that here $\ell_g$ is in a different gauge from that in our previous work, Ref.~\cite{Chen2018a}.

\begin{figure}
	\centering
	\includegraphics[width=3.5 in]{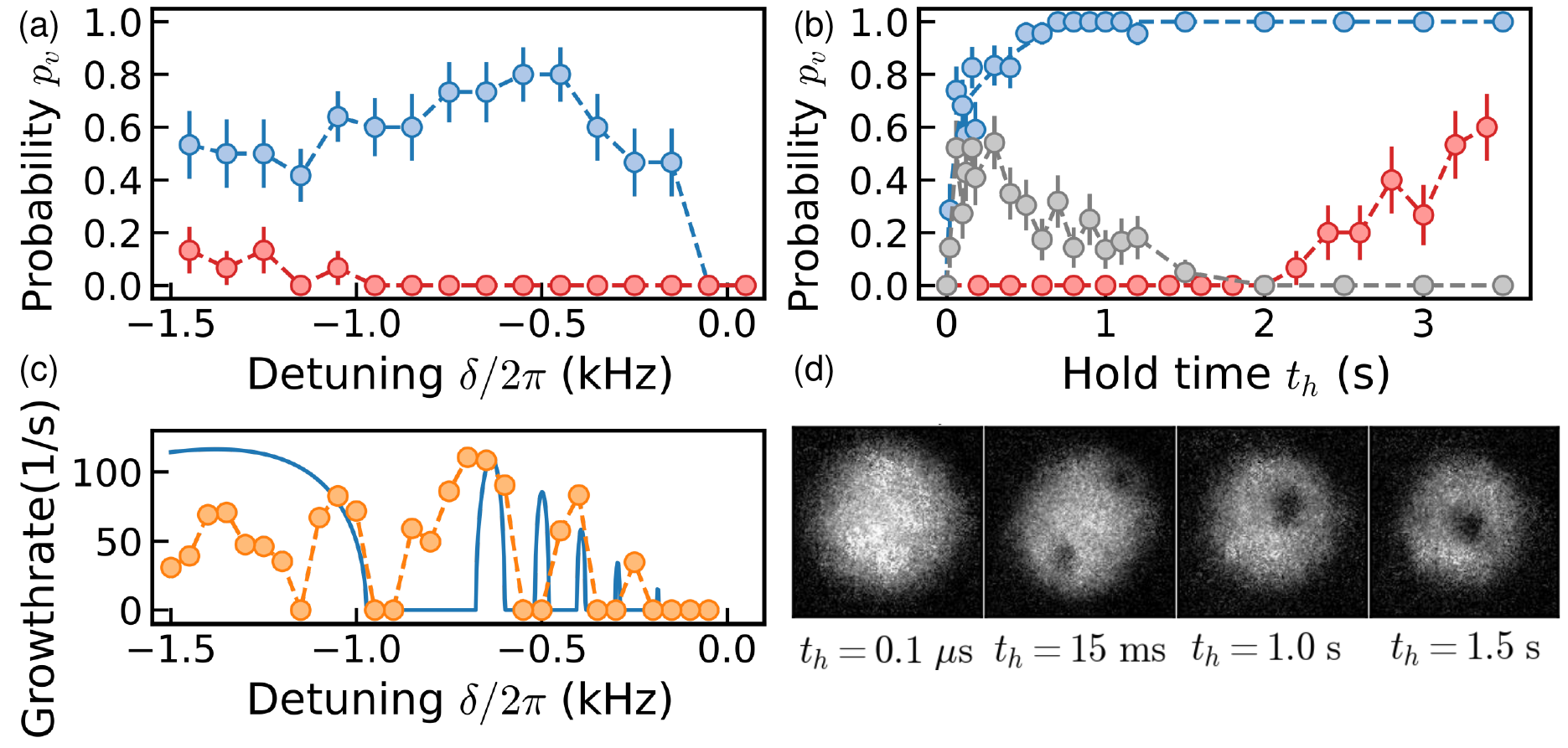}
	\caption{(a) Probability of having vortices $p_v$ vs. detuning $\delta$ at $t_h=0.1~\mu$s (red) and $0.2$~s (blue). Finite $p_v$ with $t_h=0.1~\mu$s is attributed to the loading process. (b) $p_v$ vs. $t_h$ at $\delta/2\pi=-600$~Hz (blue) and $100$~Hz (red). Gray symbols show probability of having more than one vortex $p_{v,N_v>1}$ at $\delta/2\pi = -600$~Hz. (c) Growth rates of the unstable mode in a circularly symmetric system obtained as imaginary parts of the 2D Bogoliubov spectrum (blue) and by 3D TDGPE simulation (orange). (d) Atomic optical density vs. $(x,y)$ for $\delta/2\pi= -600$Hz. For $t_h=1.0$ and $1.5$~s, the representative images are both single-vortex, while there are small but nonzero probabilities of observing more than one vortex, $p_{v,N_v>1}$, as shown in Fig.~2b. The field of view is $175 \times 175 \micron^2$.}
	\label{Fig:pv}
\end{figure}

Our experiment starts with a BEC in $|m_F=-1\rangle$ in a crossed dipole trap with $N\approx~ 1.35\times 10^5$ atoms. The trap frequencies along the $x,y,z$ directions are $(\omega_x,\omega_y,\omega_z)/2\pi=$(120,120,157)~Hz, which gives the radial Thomas-Fermi radius $R_{\rm TF}\approx 6.7\micron$.  We load the atoms into $\varphi\ket{\xi_{-1}}$ by turning on the Raman coupling in 7 ms and then ramping the Raman detuning from $\delta_i=\delta+2\pi\times 2600$~Hz to $\delta$ with the rate $d\delta/d t=-2\pi \times 178.6$~Hz/ms. The value of $\delta$ tunes the implemented $A_{-1}(r)$. We hold the system at $\delta$ for $t_h$.  Due to heating from the Raman coupling during $t_h$, atoms can populate the excited states $|\xi_{0,1}\rangle$.  To eliminate the excited atoms and probe $\varphi$ associated with $|\xi_{-1}\rangle$, we perform ``deloading''~\cite{Chen2018a,Williams2012}, here being an inverse of the loading process, where we adiabatically turn off $\vec A$.  This maps the dressed state $|\xi_n\rangle$ to the bare spin state $|m_F=n\rangle$, after which $\varphi$ is obtained by selectively imaging $|m_F=-1\rangle$ component after a 23.9~ms time-of-flight in all measurements~\footnote{Spin-resolved imaging is not practically useful in our experiment due to technical reasons. The signal-to-noise ratio of individual $\ket{m_F}$ component in the imaging is insufficient during the $t_h$ owing to two factors: the minor $\ket{m_F=-1}$ has small OD in the detuning range of interest and the thermal atoms populating $\ket{\xi_{0,1}}$. See supplemental material.}.

We note that the spinor wave function during the loading and deloading processes are still expressed as $\varphi|\xi_{-1}\rangle$ with time-dependent $|\xi_{-1}\rangle$ such that $|\xi_{-1}\rangle=|m_F=-1\rangle$ before loading and after deloading. Although $\varphi$ evolves during the loading/deloading, we numerically confirmed its cOAM $\tilde{L}_z$ is almost unchanged. Thus, the initial vortex-free state in $|m_F=-1\rangle$ leads to a vortex-free $\varphi$ at $t_h=0$ right after the loading. Furthermore, at any $t_h$ the observation of the density profile and the OAM $L_z$ of $|m_F=-1\rangle$ after deloading give the information on the vortex configuration and cOAM $\tilde{L}_z$ before deloading, equal to $L_z$.

We first investigate the probability of having vortices, $p_v$, with various $\delta$ and $t_h$. We repeat the experiment by 15 times at given $(\delta,t_h)$, identify for each image whether the BEC has vortices or not from the density dip signifying the phase singularity of a vortex, and derive the probability $p_v$~\footnote{The uncertainty of $p_v$ in the $n=15$ shots in Fig.~\ref{Fig:pv} is given by $\sigma/\sqrt{n}$ in the binomial distribution, where $\sigma$ is the standard deviation of the $n$ shots.}. Fig.~\ref{Fig:pv}a shows the $\delta$ dependence of $p_v$ at $t_h=0.2$~s, where vortices appear with high probability at $\delta \lesssim 0$ and $p_v$ peaks at $\delta/2\pi\sim -500$~Hz. The threshold detuning of vortex nucleation is $\delta_{\rm thr}/2\pi \sim -50$~Hz, which corresponds to the onset of instability. We note that $\delta_{\rm thr}/2\pi$ is significantly below 200~Hz, which is the lower critical detuning for thermodynamically stable $\ell=0$ state. This is the same observation as that in the aforementioned mechanical stirring BECs~\cite{Madison00,Madison01,Chevy2002}.

In Fig.~\ref{Fig:pv}b, we plot the $t_h$ dependence of $p_v$ at $\delta/2\pi=-600$ and 100 Hz, together with that of the probability of having more than one vortex, $p_{v, N_v>1}$, at $\delta/2\pi=-600$~Hz. At $\delta/2\pi=100$~Hz, $p_{v, N_v>0}=0$ always holds, and a vortex is nucleated only at $t_h\gtrsim 2.0$~s, which may be due to the effect of thermal atoms~\cite{Lobo2004}. On the other hand,  at $\delta/2\pi=-600$~Hz, $p_v$ starts increasing from $t_h=0$, and multiple vortices arise in the early stage. Fig.~\ref{Fig:pv}d depicts the images of the atomic optical densities (ODs) for various $t_h$ at $\delta/2\pi=-600$~Hz. One can see two density dips in the image at $t_h=15$~ms, which is the representative image for the appearance of multiple vortices. When we use a final detuning of deloading smaller than $\delta_i$, for which the configuration of the phase singular points changes less during the deloading process, the pair of density dips comes closer to the trap center (see supplemental material). Together with the measured OAM $L_z\sim 0$ and the numerical simulations (see below), we conclude that a vortex-antivortex pair is generated near the trap center.

Indeed, the linear stability analysis predicts the appearance of an unstable mode localized at the trap center. As shown in Fig.~1b, the azimuthal velocity around the trap center increases as $\delta$ decreases. When it exceeds the critical velocity, the Landau instability arises, i.e., the Bogoliubov-de Gennes (BdG) mode localized at the trap center has a negative frequency. This localized mode can become dynamically unstable when it couples with one of the positive-frequency modes. We have numerically computed the BdG spectrum for a 2D cylindrically symmetric system and confirmed the appearance of a negative frequency mode at $\delta/2\pi\lesssim -200$~Hz, which becomes complex at several values of $\delta/2\pi<-200$~Hz (Fig.~\ref{Fig:pv}c), reasonably close to $\delta_\textrm{thr}/2\pi\sim-50$~Hz. We have also confirmed the existence of dynamical instability in 3D (see supplemental material).

\begin{figure}
	\centering
	\includegraphics[width=3.5 in]{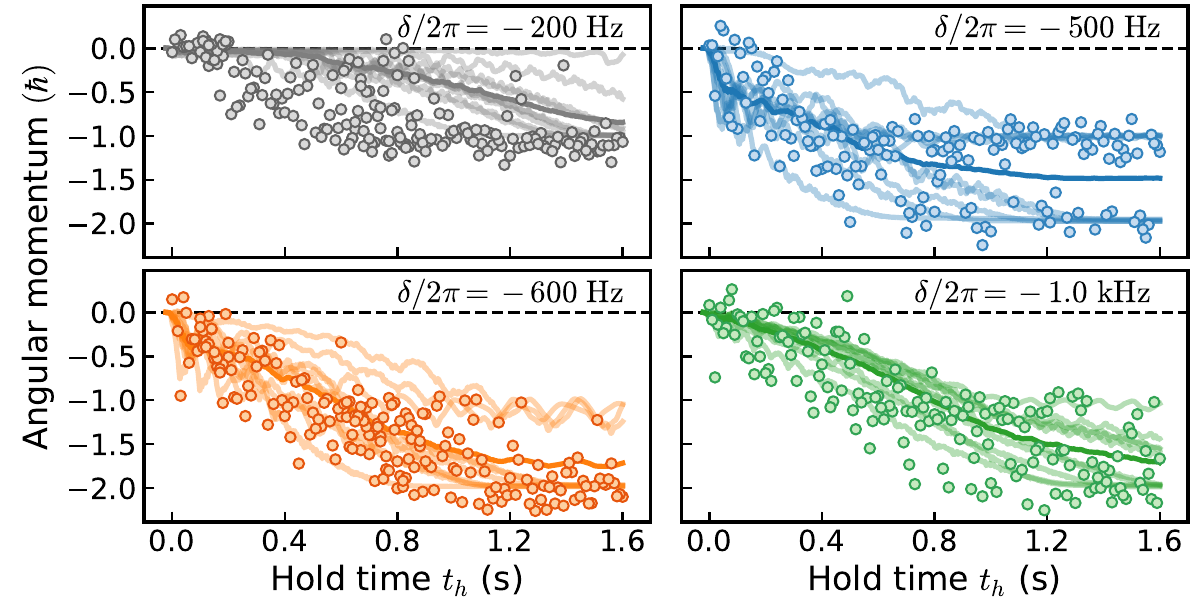}
	\caption{Angular momentum vs. $t_h$ of the atoms deloaded to $m_F=-1$ at $\delta/2\pi= -200,-500,-600, -1000$~Hz. Symbols denote the experimental data $L_z$; light-colored curves denote ten individual simulations of $\tilde{L}_z$ and the dark-colored ones indicate the average for each detuning.}
	\label{Fig:Lz vs t_h}
\end{figure}

Next, we measure the OAM $L_z$ from the quadrupole mode precession rate~\cite{Zambelli1998,Chevy2000,Riedl2009}. The quadrupole mode precession angle $\theta$ after TOF is given by
$\theta=L_z/2m R_{\bot}^2 ( \tau+\tau_{\rm exp} )$, where $L_z/2m R_{\bot}^2$ is the in-trap precession rate~\cite{Zambelli1998}, $R_{\bot}$ is the transverse size, and $\tau_{\rm exp}$ is an additional time accounting for the precession during TOF. Fig.~\ref{Fig:Lz vs t_h} illustrates the $t_h$ dependence of $L_z$ for $\delta/2\pi=-200, -500, -600$, and $-1000$~Hz. 

To quantitatively analyze the dynamics, we numerically solve the 3D 3-component Gross-Pitaevskii equation (TDGPE) and calculate the time evolution of $\tilde{L}_z$, as shown in Fig.~\ref{Fig:Lz vs t_h}. The TDGPE starts from an initial spin-polarized state before the loading process. We add a small noise in the initial state and incorporate the cylindrical asymmetry of the Raman coupling. The latter is owing to imperfect optical alignments and implemented by adding asymmetry terms with amplitudes experimentally measured and relative phases randomly chosen in each run. We also introduce phenomenological energy dissipation, which is necessary for the system to reach the ground-state OAM. The energy dissipation is attributed to the thermal components and is stronger for smaller $|\delta|$ and longer $t_h$~\cite{Chen2018a}. However, we use a constant dissipation to reduce the number of unknown parameters in the simulation, where in Fig.~\ref{Fig:Lz vs t_h} the magnitude of the dissipation is determined so as to agree with the experimental data at $\delta/2\pi=-600$ Hz. The numerical data reasonably agrees with the experiment also for $\delta/2\pi=-500$ and $-1000$~Hz, but the numerical result for $\delta/2\pi=-200$~Hz proceeds slower than the experiment, consistent with a $\delta$-dependent dissipation.

\begin{figure}
	\centering
	\includegraphics[width=3.5 in]{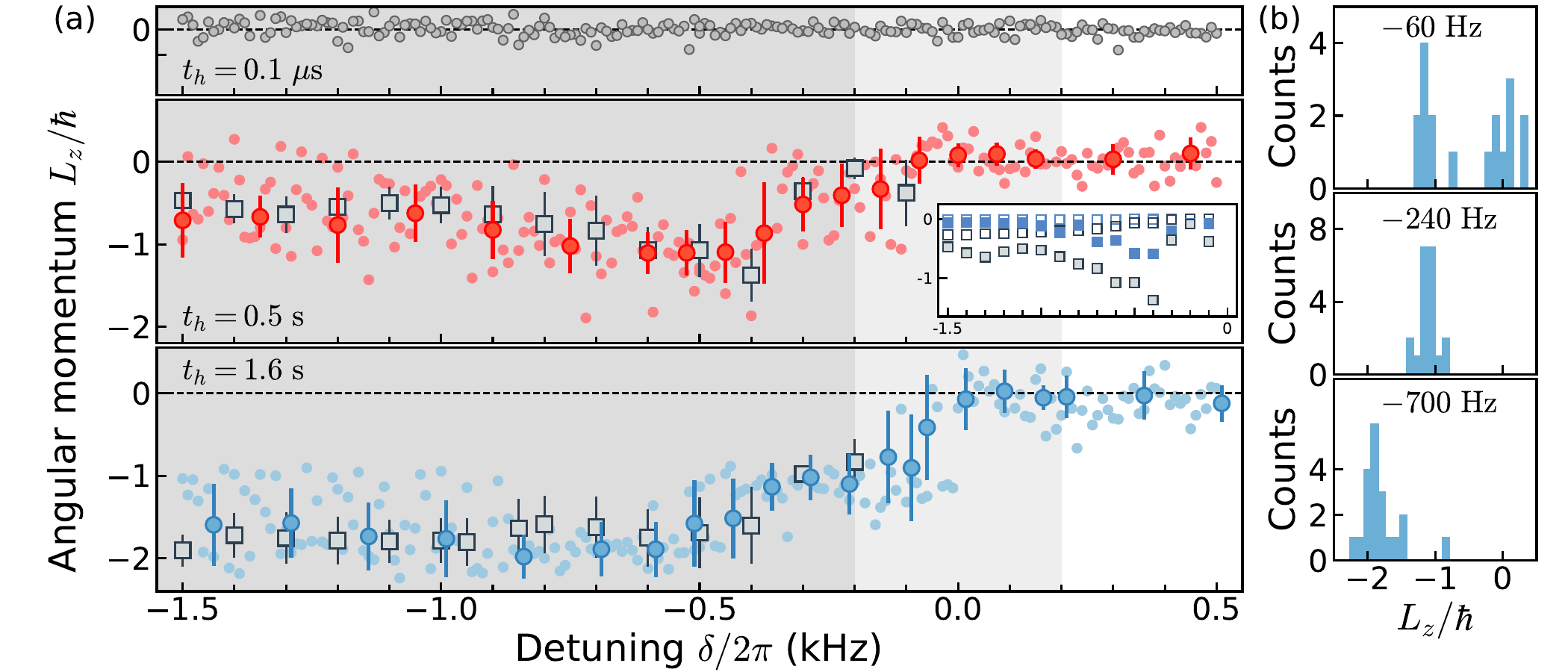}
	\caption{(a) Angular momentum $L_z$ of atoms deloaded to $m_F= -1$ versus detuning at various hold time, $t_h=0.1~\mu$s, $0.5$~s and $1.6$~s. The data with a negligibly small $t_h=0.1~\mu$s indicate that $L_z$ remains about zero during the loading and deloading process. Average and standard deviation of 15 points taken at each $\delta$ are also displayed for $t_h=0.5$~s and $1.6$~s. Squares display the numerical data $\tilde{L}_z$ obtained from 10(30) simulations for $t_h=0.5(1.6)$~s. The background colors indicate the ground state phases of $\ell_g=0$ (white), $-\hbar$ (light gray) and $-2\hbar$ (gray). The inset of the middle panel shows $\tilde{L}_z(\delta)$; the light-gray (blue)-filled symbols denote with asymmetry and with (without) dissipation and the black (blue) open symbols denote without asymmetry and with (without) dissipation. (b) Histograms of $L_z$ at $\delta/2\pi= -60, -240, -700$~Hz and $t_h=1.6$~s. 
	}
	\label{Fig:Lz vs detuning}
\end{figure}

We also measure the $\delta$ dependence of $L_z$ for $t_h=0.1\mu, 0.5, 1.6$~s in Fig.~\ref{Fig:Lz vs detuning}a. First, we compare the result for $t_h=0.5$~s with Fig.~\ref{Fig:pv}a and see that the short-hold-time $|L_z|$ well captures the $\delta$-dependence of $p_v$, i.e., both become nonzero at $\delta<\delta_\textrm{thr}\sim 0$ and have a peak at $\delta/2\pi\sim -500$~Hz. The peak is due to the asymmetry: We numerically confirmed that the peak disappears when no asymmetry is introduced. On the other hand, in the presence of asymmetry, the same peak appears even without the energy dissipation (Fig.~\ref{Fig:Lz vs detuning}a inset), indicating that the dissipation is not necessary to reproduce observed $L_z(\delta,t_h=0.5$~s) in Fig.~\ref{Fig:Lz vs detuning}a under Landau instability. We thus conclude that vortex nucleation results from dynamical instabilities. Consider the deviation of the $L_z(\delta)$ from the 3D growth rate for symmetric systems in Fig.~\ref{Fig:pv}c: our actual asymmetric system with a shifted LG beam center from the BEC center is expected to enhance the region for dynamical instability. We also note that the transverse component of $\vec{\Omega}_\textrm{eff}$ becomes prominent for $|\delta|\ll \Omega(R_{\rm TF})\sim 2\pi \times 1.5$~kHz, so the asymmetry in $\Omega(r)$ disturbs the system more for smaller $|\delta|$, resulting in a single peak in $L_z$ at $\delta/2\pi\sim -500$~Hz.

\begin{figure}
	\centering
	\includegraphics[width=3.5 in]{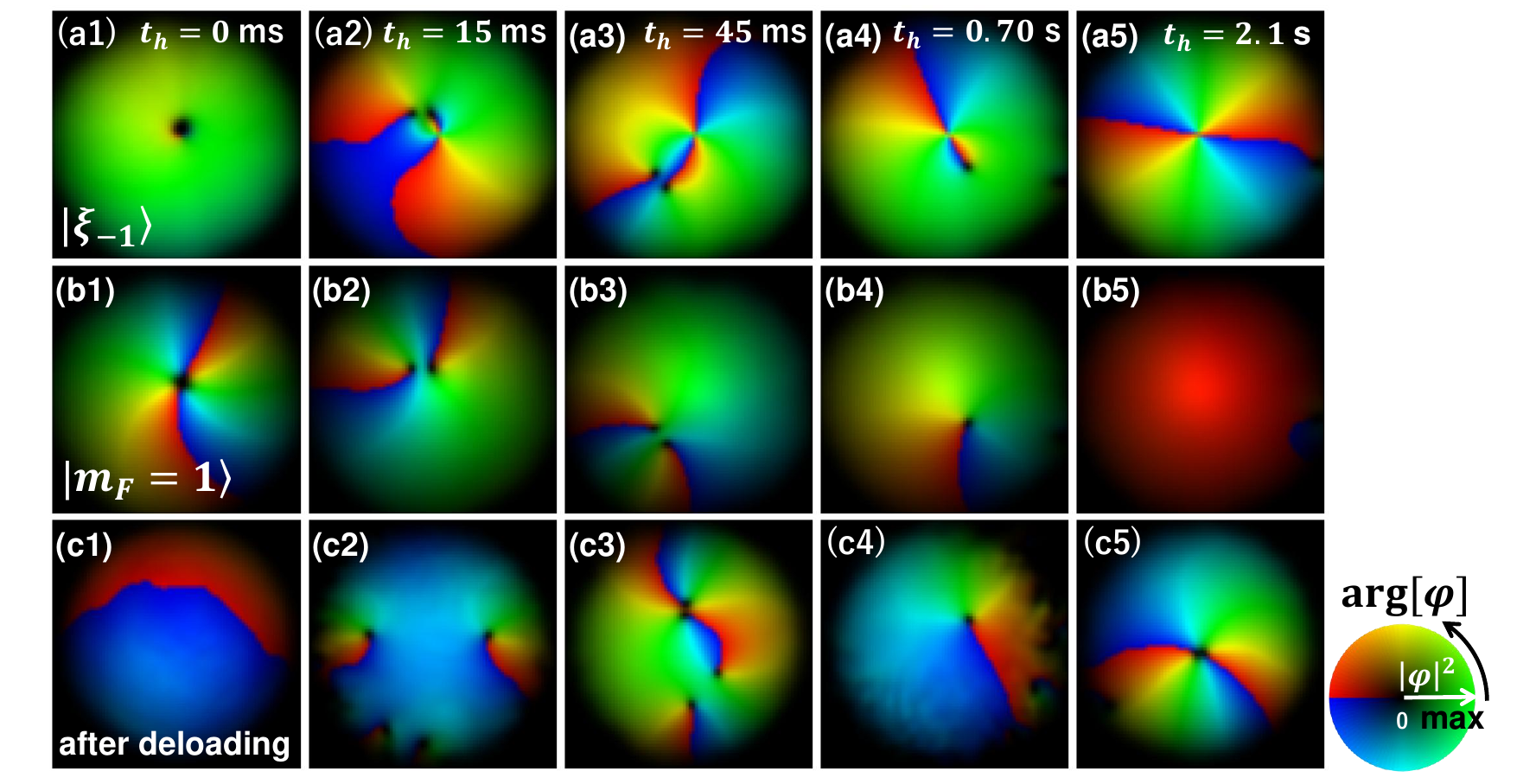}
	\caption{(a-c) Simulations of in-situ wave functions on the $z=0$ plane projected onto $|\xi_{-1}\rangle$ (a) and $\ket{m_F=1}$ (b), and the one in $\ket{m_F=-1}$ after deloading (c); the field of view is $14 \times 14 \micron^2$. Shown are the snapshots in a single run at $\delta/2\pi=-600$~Hz, where saturation indicates the atomic density normalized by the maximum value in each panel, and the hue shows the phase. The description in this paper is based on (a). At $t_h=0$, $\varphi$ is vortex-free with $w=0$ (a1). Vortex-antivortex pairs are generated near the LG beam center (a2); With the aid of asymmetry, the vortex and antivortex move apart (a3), and the system emits one (a4) or two (a5) $w=1$ vortices, enabling $L_z$ to reach $-\hbar$ or $-2\hbar$, respectively. The profiles in (b) are approximately equivalent to adding $\Delta w = 2$ at the center of the LG beam to the wave function $\varphi$ shown in (a), due to a change of basis (or gauge) between $\ket{m_F = 1}$ and $\ket{\xi_{-1}}$.  A doubly quantized vortex initially imprinted at the LG beam center (b1) splits into two (b2), which shift from the LG beam center with the aid of asymmetry (b3) and escape from the condensate one by one (b4, b5). In the experiment, we observe the wave function after deloading (c) instead of (a). The number of phase singularities and cOAM in (a) is preserved after deloading in (c), while vortex-antivortex pair annihilation occurs during deloading depending on the location of vortices before deloading. For example, at $t_h=15$~ms, although there are two vortex-antivortex pairs in $\ket{\xi_{-1}}$ (a2), one pair annihilates during deloading, leaving single pair in the deloaded $\ket{m_F=-1}$ (c2).
	}
	\label{Fig:simulations}
\end{figure}

Next, we discuss the late-time dynamics. The data at $t_h=1.6$~s has plateaus at $0, -\hbar, -2\hbar$ where the standard deviations of $L_z$ are relatively small, $\approx 0.2 \hbar$, and are close to the typical uncertainty of $L_z$ for the case when BECs have stable $L_z$. For $-400$ Hz $\lesssim \delta/2\pi\lesssim -200$~Hz, $L_z$ remains $-\hbar$, suggesting the existence of the metastable state.	This can be understood as follows: To reach the ground state with $L_z=\ell_g=-2\hbar$, two phase winding $w=1$ vortices among the pair-created vortices have to move out from the condensate (see also Fig.~\ref{Fig:simulations}ac);	Because of the $r$-dependent gauge potential $\vec{A}$, even when $|\vec{A}|$ at the cloud center is large enough to introduce the instability, $|\vec{A}|$ in the whole condensate can be insufficient to emit the two vortices from the condensate, resulting in an energetically metastable $L_z=-\hbar$ state (see supplemental material). When $|\vec{A}(r)|$ becomes large enough, $L_z$ can reach $-2\hbar$ (see the bottom panel of Fig.~\ref{Fig:Lz vs detuning}a at $-850$ Hz $\lesssim \delta/2\pi\lesssim -550$~Hz).
For the detunings at transitions between the plateaus, the standard deviations of $L_z$ are relatively large, correspondingly, the histogram of $L_z$ has two peaks (see the top panel of Fig.~\ref{Fig:Lz vs detuning}b). This behavior is also shown in the $\delta/2\pi=-500$ Hz data of Fig.~\ref{Fig:Lz vs t_h}. For $\delta/2\pi \lesssim -900$~Hz with $t_h=1.6$~s, $|L_z|$ decreases to $<2\hbar$ and the standard deviation becomes larger, which is likely due to reduced thermal atom fraction in $\ket{\xi_{0,1}}$, and thus reduced dissipation, at $|\delta/2\pi| \lesssim 1$~kHz (see supplemental material and~\cite{Chen2018a}). The thermal atom fraction depends also on $t_h$. However, our simulation employs a constant dissipation that is independent of $\delta$ and $t_h$, which can explain the deviation between the simulations and experimental data of $t_h=1.6$~s.

Finally, we comment on the relation with the splitting dynamics of a doubly quantized vortex~\cite{Shin2004,Moon2015,Weiss2019}. When we focus on the bare spin $|m_F=1\rangle$ component, a doubly quantized vortex is imprinted at $t_h=0$, which is dynamically unstable against splitting in the limit of $\delta\to-\infty$~\cite{Pu1999,Mottonen2003}. A similar instability exists even at a small negative $\delta$, and thus we can capture the dynamics in our system as vortex splitting dynamics by choosing a different gauge for $|\xi_{-1}\rangle$. In this view,  the gauge potential makes it more difficult for the split vortices to leave the condensate and delays the dynamics. See $L_z$ vs. $\delta$ at $t_h=1.6$~s in Fig.~\ref{Fig:Lz vs detuning}a. Here, we stress again that we are interested in the vortex nucleation dynamics starting from a vortex-free wave function under a nonuniform synthetic magnetic field; Thus we describe our system using $|\xi_{-1}\rangle$, which is a spinor eigenstate under the SOAMC, rather than $|m_F=1\rangle$. We summarize the correspondence between the dynamics of the wave functions in $|\xi_{-1}\rangle$, $|m_F=1\rangle$, and that after deloading in Fig.~\ref{Fig:simulations}.

In conclusion, we observe vortex nucleations in spinor BECs  which are initiated by a spatially localized unstable mode, owing to an azimuthal velocity fields that peaks near the trap center. A vortex-antivortex pair creation near the center signifies the dynamically unstable mode that leads to vortex nucleations. The experimental data is consistent with numerical simulations. We present the first experimental characterization of OAM's time evolution during vortex nucleations. We may extend the current work to more versatile vortex configurations with dynamical manipulations and higher order Raman vortex laser beams with $\Delta \ell/\hbar>1$. 
Our calculations show that one may produce instability at $\delta > 0$ with sufficiently large $\Delta \ell$. The location of the peak of the velocity field at $r=r_{\rm max}$ can be engineered, and one expects unstable surface modes for system size $R < r_{\rm max}$ and localized modes for $R > r_{\rm max}$, and intriguing competitions between these two mechanisms.

\begin{acknowledgments}
The authors thank C. Chin,  W.~D. Phillips, Jean Dalibard and I.~B. Spielman for useful discussions. We also thank N.~C. Chiu for critical readings of the manuscript, and thank H.~C. Yao, T.~H. Chien and Y.~H. Su for their technical assistance. Y.~-J.~L. was supported by NSTC 108-2112-M-001-033-MY3 and 111-2112-M-001-048-MY3 and the Thematic Research Program of Academia Sinica
S.~-K.~Y. was supported by NSTC 110-2112-M-001-051-MY3. Y. K. was supported by JSPS KAKENHI (Grant Nos. JP19H01824, JP23K20817, and JP24K00557).
\end{acknowledgments}


\widetext
\clearpage
\begin{center}
	\textbf{\large Supplemental Material: Vortex nucleations in spinor Bose condensates under localized synthetic magnetic fields}
\end{center}
\setcounter{equation}{0}
\setcounter{figure}{0}
\setcounter{table}{0}
\setcounter{page}{1}
\makeatletter
\renewcommand{\theequation}{S\arabic{equation}}
\renewcommand{\thefigure}{S\arabic{figure}}
\renewcommand{\bibnumfmt}[1]{[S#1]}
\renewcommand{\citenumfont}[1]{S#1}

\maketitle

\section{Formalism of the dressed states and associated gauge potentials}
The Hamiltonian in the bare spin basis, $\ket{m_F=1,0,-1}$, in the frame rotating at $\Delta
\omega_L$ under rotating wave approximation in the $(r,\phi,z)$ coordinate is
\begin{align}\label{eq:Hlab}
	\hat{H}_{\rm lab}=\left[\frac{-\hbar^2}{2m}\frac{\partial}{r\partial
		r}(r \frac{\partial}{\partial
		r})-\frac{\hbar^2}{2m}\frac{\partial^2}{\partial z^2}
	+\frac{\hat{\ell}^2}{2m r^2}\right] \otimes {\hat 1}+\vec{\Omega}_{\rm
		eff}\cdot \vec{F}-\hbar \omega_q 
	\left({\hat 1}-\hbar^{-2}F_z^2 \right),
\end{align}
where $\Delta\omega_L$ is the frequency difference between two Raman laser beams, $F_x,F_y,F_z$ are the spin $1$ matrices, $\hat{\ell}= -i \hbar \partial_{\phi}$ is the orbital angular momentum (OAM) operator, and $\omega_q$ is the quadratic Zeeman shift. The effect of the small $\omega_q/2\pi=50$~Hz in negligible in our experiment. Here, the effective Zeeman field from the Raman beams is $\vec{\Omega}_{\rm eff}=\Omega(r)\cos \phi{\mathbf e}_{x}-\Omega(r)\sin \phi{\mathbf
	e}_{y}+\delta{\mathbf e}_{z}$ given by the spin-OAM coupling (SOAMC)~\cite{SChen2018} where the OAM transfer is $\Delta\ell=\hbar$ between $\ket{m_F}$ and $\ket{m_F+1}$, $\delta=\Delta\omega_L-\omega_Z$ is the Raman detuning, and $\omega_Z$ is the linear Zeeman shift. SOAMC can be employed to create topological excitations~\cite{SChen2018}, analogous to those created by spin rotation methods using magnetic fields~\cite{Leanhardt2003,Choi2012a,Ray2014,SWeiss2019,Xiao2021}.

For sufficiently large $\vec{\Omega}_{\rm eff}\cdot
\vec{F}$, the motional kinetic energy $-(\hbar^2/2m)
\nabla^2$ of the atoms is negligible and the energy eigenstates of the overall
Hamiltonian are well approximated by the eigenstates of $\vec{\Omega}_{\rm eff}\cdot \vec{F}$, $|\xi_n\rangle$. Here, we label the lowest-, middle-, and highest-energy dressed states as $|\xi_{-1}\rangle, |\xi_0\rangle$, and $|\xi_1\rangle$, respectively, which are normalized at each $\vec{r}$, i.e., $\langle\xi_n(\vec{r})|\xi_n(\vec{r})\rangle=1$. Under this approximation,
the atom's spinor wave function follows the local dressed eigenstate
$|\xi_n\rangle$, whose quantization axis is along $\vec{\Omega}_{\rm eff}$.
When all atoms are in the lowest-energy dressed state, the condensate wave function is described by $|\Psi(\vec{r})\rangle=\varphi(\vec{r})|\xi_{-1}\rangle$, where $\varphi$ is the external part of the wave function whose amplitude square giving the condensate density. The effective Hamiltonian for the external wave function $\varphi$ is~\cite{SChen2018,SChen2018a}
\begin{eqnarray}
	H_{\rm eff}=\frac{-\hbar^2}{2m}\frac{\partial}{r\partial
		r}(r \frac{\partial}{\partial
		r})-\frac{\hbar^2}{2m}\frac{\partial^2}{\partial z^2}
	+\frac{\left(\hat{\ell}-r A_{-1}\right)^2}{2m r^2} + V(r)+\varepsilon_{-1}+W_{-1}.
	\label{eq:projectedH}
\end{eqnarray}
Here $\hat{\ell}$ is the canonical angular momentum
operator when it operates on $\varphi$, and $\vec{A}=i\hbar\langle\xi_{-1}|\vec{\nabla}|\xi_{-1}\rangle=A_{-1}\hat{\phi}$ where $A_{-1}(r)=
(i\hbar/r)\langle \xi_{-1}|\partial_{\phi} \xi_{-1}\rangle$ is the azimuthal gauge potential. $V(r)$ is the spin-independent trap,
$\varepsilon_{-1}= -\sqrt{\Omega(r)^2+\delta^2}$ is the eigenenergy of
$\vec{\Omega}_{\rm eff}\cdot \vec{F}$, and $W_{-1}\approx \hbar^2/2mr^2$ is the geometric scalar potential.

In this paper, we load the atoms to the lowest-energy dressed state $\ket{\xi_{-1}}$. A general $|\xi_{-1}\rangle$ is given by Euler rotations~\cite{Ho1998} as
\begin{eqnarray}
	|\xi_{-1}\rangle &= e^{i (\bar{\theta}+\bar{\gamma})}\left( e^{i \phi}\frac{1 -
		\cos \beta}{2}, -\frac{\sin \beta}{\sqrt{2}}, e^{-i \phi}\frac{1 +
		\cos \beta}{2} \right)^{\rm T}, \rm {general},
\end{eqnarray}
where $\beta(r)=\tan^{-1} [ \Omega(r)/\delta]$ is the polar angle of
$\vec{\Omega}_{\rm eff}$, and $\bar{\theta}+\bar{\gamma}$ is the phase for gauge transformation. In this paper, we choose $\bar{\theta}+\bar{\gamma}$ such that the initial external wave function $\varphi$ is the eigenstate of $\hat{\ell}$ with eigenvalue
$\ell=0$, i.e., vortex-free. In our experiment, we start with a BEC spin-polarized in $\ket{m_F=-1}$ with zero OAM and load it into $\ket{\xi_{-1}}$. Consequently, the initial dressed state $\ket{\Psi(t_h=0)}$ composes bare spin $|m_F=1,0,-1\rangle$ components with OAM of $2\hbar,\hbar,0\hbar$, respectively, where the $\ket{m_F=-1}$'s OAM remains zero, same as that before the dressed state loading. Thus, we choose $\bar{\theta}+\bar{\gamma}=\phi$, and define $\ket{\xi_{-1}}$ as
\begin{align}
	|\xi_{-1}\rangle &= \left( e^{i 2\phi}\frac{1 -
		\cos \beta}{2}, -e^{i \phi}\frac{\sin \beta}{\sqrt{2}},\frac{1 +
		\cos \beta}{2} \right)^{\rm T} {\rm with~} \bar{\theta}+\bar{\gamma}=\phi.
	\label{eqn:xi_minus1}
\end{align}
Under this gauge choice, the bare spin $|m_F=1,0,-1\rangle$ components in a dressed state $\varphi|\xi_{-1}\rangle$ has the OAM $\ell+2\hbar,\ell+\hbar,\ell$, respectively.

Because the synthetic gauge field corresponding to $|\xi_{-1}\rangle$ in Eq.~\eqref{eqn:xi_minus1} is given by
\begin{align}
	A_{-1}&=\frac{\hbar}{r}(\cos \beta - 1)
	=\frac{\hbar}{r}\left[\frac{\delta}{(\Omega(r)^2+\delta^2)^{1/2}}-1\right],
	\label{eqn:gauge_Aplus}
\end{align}
the azimuthal velocity of the condensate for the case when $\varphi$ has the phase winding number $\ell/\hbar$ is given by
\begin{align}
	\vec{v}_{-1}&=m^{-1}(\hbar \vec{\nabla} \tilde{\vartheta}-A_{-1}\hat{\phi})
	=m^{-1}\left(\frac{\ell}{r} -A_{-1}\right)\hat{\phi},\nonumber\\
	v_{-1}&=m^{-1}\left(\frac{\ell}{r} -A_{-1}\right).
	\label{eqn:velocity}
\end{align}

An alternative gauge $\bar{\theta}+\bar{\gamma}=0$ can be used, such as in our previous work, Ref.~\cite{SChen2018a}, leading to
\begin{eqnarray}
	A_{-1}^0=\frac{\hbar}{r}\cos \beta.
\end{eqnarray}
Let $\tilde{\ell}$ denote the angular momentum of $\varphi$ in this gauge. Then, $\tilde{\ell}+\hbar,\tilde{\ell},\tilde{\ell}-\hbar$ are the OAM of
the bare spin $|m_F=1,0,-1\rangle$ components of the dressed state
$\varphi|\xi_{-1}\rangle$, respectively. The kinetic angular momentum of the dressed state is gauge independent, i.e., $\tilde{\ell}-rA_{-1}^0=\ell-rA_{-1}$.

\begin{figure}
	\centering
	\includegraphics[width=7 in]{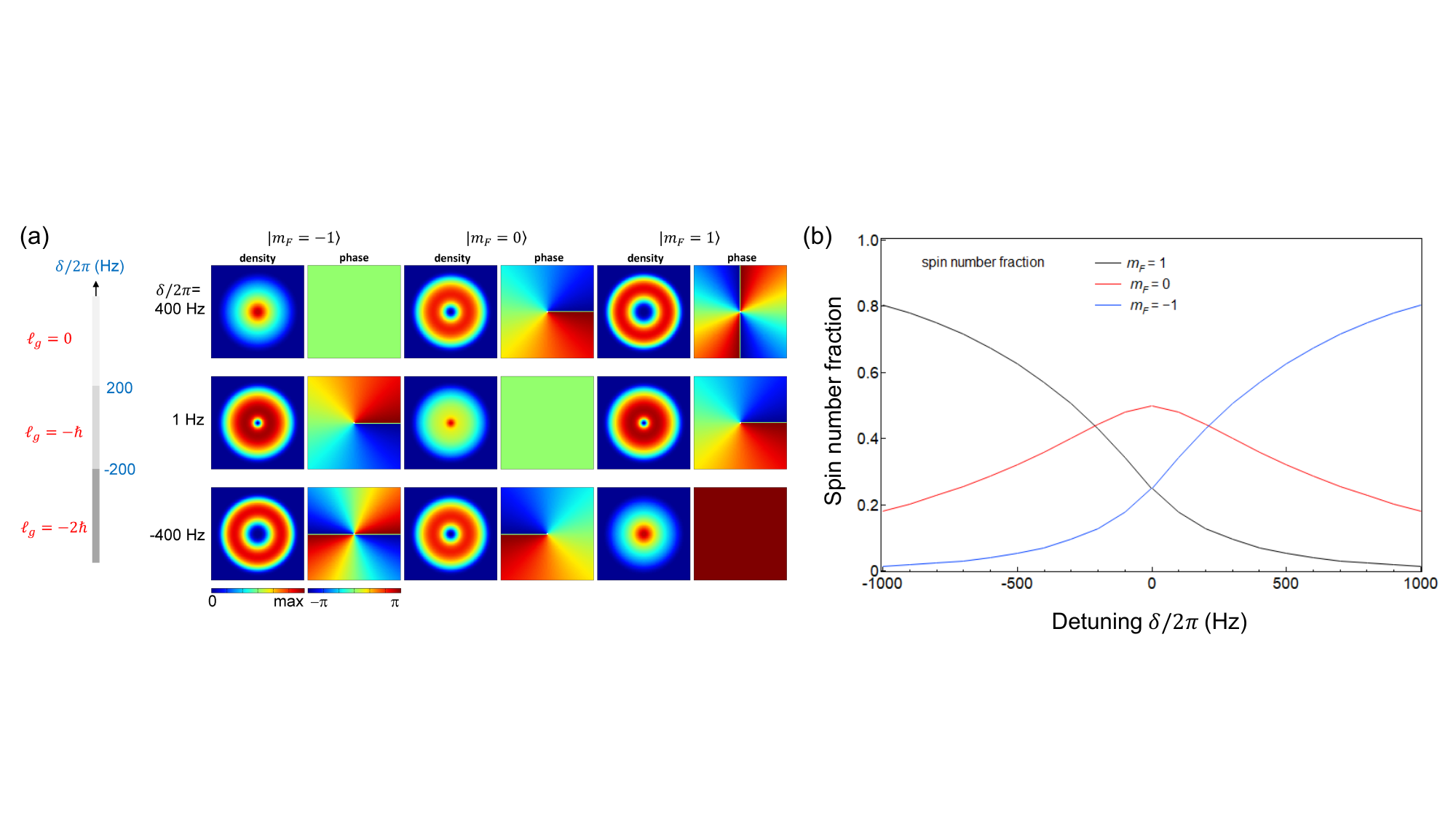}
	\caption{(a) The canonical OAM of the absolute ground state $\ell_g$ vs. detuning $\delta$ and the Gross-Pitaevskii ground state with $\ell_g= 0,-\hbar,-2\hbar$ at $\delta/2\pi=400,1,-400$~Hz in the top, middle and bottom row, respectively. (b) Spin number fraction of $\ket{m_F=\pm 1,0}$ vs. detuning $\delta$.}
	\label{fig:groundstate}
\end{figure}

The canonical OAM of the absolute ground state $\ell_g$ of the dressed state depends on detuning $\delta$~\cite{SChen2018a}, as shown in Fig.~\ref{fig:groundstate}a. This can be understood with the spin number fraction vs. detuning of $\ket{\xi_{-1}}$ in Fig.~\ref{fig:groundstate}b: For $\delta/2\pi>200$~Hz ($<-200$~Hz), the largest spinor population is in $m_F=-1(1)$. For $|\delta/2\pi|<200$~Hz, the largest spinor population is in $m_F=0$. Consider the lowest energy ground dressed state, the vortex kinetic energy is minimized. Therefore, the mechanical OAM of $m_F=-1,0,1$ should be zero for $\delta/2\pi>200$~Hz, $|\delta/2\pi|<200$~Hz, $\delta/2\pi<-200$~Hz, respectively, corresponding to the ground state’s canonical OAM (cOAM) $\ell_g= 0,-\hbar,-2\hbar$, respectively. Fig.~\ref{fig:groundstate}a also displays three examples of the dressed state images with $\ell_g= 0,-\hbar,-2\hbar$, respectively.

\section{Methods}
\subsection{Experimental procedures}
In the beginning of the experiment we produce a $^{87}$Rb BEC with $N \approx 1.35 \times 10^5$ atoms in a crossed dipole trap in $|F,m_F\rangle=|1,-1\rangle$~\cite{SChen2018a}. The trap frequencies along ${\mathbf e}_{x},{\mathbf e}_{y},{\mathbf e}_{z}$ directions are $(\omega_x,\omega_y,\omega_z)/2\pi=$(120,120,157)~Hz. The smallest trap ellipticity $\epsilon=(\omega_x^2-\omega_y^2)/(\omega_x^2+\omega_y^2)$ that we can reach is typically $<0.006$.
%
%
Then we adiabatically load the $|m_F=-1\rangle$ BEC in the lowest energy Raman-dressed state, where the fraction in the excited dressed states are negligible. One of the two Raman beams is Gaussian (G) and the other one is Laguerre-Gaussian (LG). The beams are at $\lambda=790$~nm where their scalar light shifts from the D1 and D2 lines cancel. The G Raman beam has a waist $\approx200~\mu$m, and the LG Raman produced by a vortex phase plate has a phase winding number $\Delta \ell/\hbar=1$ and radial index of $0$. The G and LG beams have frequencies of $\omega_L$ and $\omega_L+\Delta \omega_L$ and are linearly polarized along ${\mathbf e}_{y}$ and ${\mathbf e}_{x}$, respectively. The Raman detuning is $\delta=\Delta\omega_L-\omega_Z$, and $\omega_Z/2\pi \approx 0.55$~MHz is the linear Zeeman shift. We estimate that the uncertainty of the relative position of the LG beam center $O^{'}$ to the BEC center $O$ is $\lesssim 0.4\micron$.

We measure the angular momentum $L_z$ of the atoms deloaded to $\ket{m_F=-1}$ as the following: right after the deloading, we excite the surface quadrupole mode by abruptly changing the trap frequencies along the $x^{\prime},y^{\prime}$ direction to $120\sqrt{0.6},120\sqrt{1.6}$~Hz, respectively, where $(x^{\prime},y^{\prime})$ has a $45$ degree angle relative to $(x,y)$. This suddenly deforms the atoms to $\epsilon^{\prime}=(1.6-0.6)/(1.6+0.6)\approx 0.4545$. We hold it at $\epsilon^{\prime}$ for $0.4$~ms, and then suddenly change the trap frequencies back to $120$~Hz in both directions, after which the quadrupole mode precesses within a delay time $\tau$ up to 9~ms; finally the atoms are released for a 23.9~ms TOF. The typical value of $\tau$ is $8.2$~ms. 

We adopt the feed-forward method to stabilize the magnetic bias field using fluxgate field sensors. After the BEC preparation we wait for the external trigger from the 60 Hz line, after which we apply feed-forward current signals into bias coils to cancel the field noise from 60 Hz harmonics. We also compensate the drifts of the DC magnetic field from the ambient and bias coils by measuring the field at the end of each experimental cycle and applying the feed-forward signal in the next cycle. Our typical field uncertainty is $\lesssim 70$~Hz $=0.1$~mG.

We compute the vortex nucleation probability as $p_v=n_v/15$ where $n_v$ is the number of experimental realizations with one or more than one vortices within the total of 15 realizations (the number of having no vortex is $15-n_v$). We identify vortices from the density dips signifying the phase singularity of a vortex. The vortex counting algorithm is based on Ref.~\cite{SAboshaeer01,SPrice2016} and references therein. We use a microwave field to selectively pump the atoms from $\ket{F=1,m_F}$ to $\ket{F=2}$ and perform resonant absorption imaging of $F=2 \rightarrow F^{\prime}=3$ transition. All the images shown in the paper are single-shot.

\subsection{Probe after deloading}
The adiabatic deloading is to map the lowest, middle, and highest energy dressed state $\ket{\xi_{-1,0,1}}$ to the single spin $\ket{m_F= -1,0,1}$, respectively. Here, our deloading procedure is the inverse of loading, i.e., ramping the detuning from $\delta$ to $\delta_{\rm del}=\delta + 2\pi\times 2600$~Hz with $d\delta/d t=2\pi \times 178.6$~Hz/ms (same $|d\delta/d t|$ as the loading) followed by turning off the Raman coupling in 7~ms.

Our main purpose of probing the atoms after deloading is to observe the wave function $\varphi$ in the spinor basis $|\xi_{-1}\rangle$, where the gauge of $|\xi_{-1}\rangle$ is chosen such that $\varphi$ is initially vortex-free with zero canonical OAM. We can also exclude atoms populating $|\xi_{0,1}\rangle$, which are thermally excited due to the heating from the Raman coupling during $t_h$ (see the next subsection ``Characterization of thermal atoms'' and Ref.\cite{SChen2018a}): we probe only the $|\xi_{-1}\rangle$ component by selectively imaging $|m_F=-1\rangle$ after a 23.9 ms time-of-flight in all measurements.

The deloading can avoid another technical issue: For increasing $t_h$ when vortex nucleations occur, 
the atomic optical density (OD) reduces such that the signal-to-noise-ratio (SNR) is low in the three individual spin images (using the standard Stern-Gerlach method). As we deload the atoms to the single $m_F=-1$, the atomic OD and SNR is sufficient for us to probe up to $t_h=1.6$~s for the precession of the quadrupole mode and up to $t_h=3.5$~s for counting vortices.

We have determined the deloading process from the numerical simulation such that the OAM $\tilde{L}_z$ of the projected wave function onto $\ket{\xi_{-1}}$ state remains almost the same as that before deloading. The adiabatic sweeping speed is important, while the final detuning $\delta_{\rm del}$ is not, because the final $\tilde{L}_z$ is almost unchanged when $\delta_{\rm del}$ is larger than a certain positive value ($\delta_{\rm del}\geq 2\pi\times 600$~Hz for the case of $\delta=2\pi\times-600$~Hz). For comparison of a standard $\delta_{\rm del}$ and a smaller $\delta_{\rm del}$, see the subsection ``Initial deformation at the onset of vortex nucleations'', Fig.~\ref{fig:deloading_asym_simu} and Fig.~\ref{fig:initial_deform_exp}. On the other hand, for turning off the Raman coupling adiabatically at the detuning $\delta_{\rm del}$ within
7 ms, $\delta_{\rm del}$ needs to be larger than $\sim 2\pi\times 1000$~Hz. We finally choose $\delta_{\rm del}=\delta+2\pi\times 2600$~Hz so that we can use the same sequence with the same deloading period for the detuning range ($\delta_{\rm del}\geq 2\pi\times -1500$~Hz for our technical convenience. We also confirm experimentally that the difference in the atoms’ $L_z$ measured with $\delta_{\rm del}=2\pi\times 2000$~Hz and a smaller final detuning $600$~Hz is smaller than the uncertainty of $L_z$. Further, we experimentally measured $L_z$ with the shortest adiabatic deloading time and with the longer deloading time that is used in our typical procedure, respectively. We verify that the difference in $L_z$ measured in the two time sequences is smaller than the uncertainty of $L_z$, which is $\approx 0.2\hbar$.

\begin{figure}
	\centering
	\includegraphics[width=6.3 in]{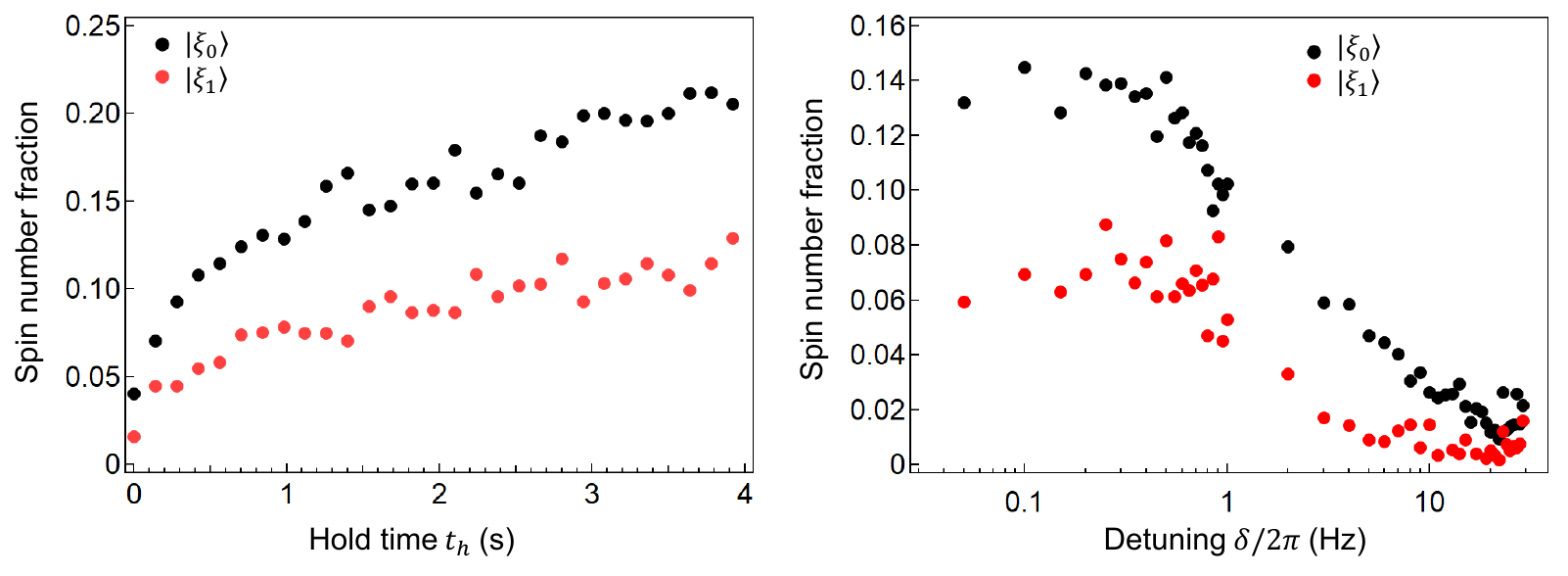}
	\caption{Number fractions of atoms in the excited dressed states $\ket{\xi_0}$ (black) and $\ket{\xi_1}$ (red) during vortex nucleations vs. hold time $t_h$ with detuning $\delta=0$ (a) and vs. $\delta$ with $t_h=1.0$~s. The atoms in $\ket{\xi_{0,1}}$ are purely thermal and accounts for an energy dissipation in the system.}
	\label{fig:thermal}
\end{figure}

\subsection{Characterization of thermal atoms}

We observe thermal atom numbers increase as the dressed state hold time $t_h$ increases~\cite{SChen2018a}. The origin is the following: At $t_h=0$, all the atoms are prepared in the lowest energy spinor branch dressed state $\ket{\xi_{-1}}$. As $t_h$ increases, the atoms start populating the excited dressed states, $\ket{\xi_{0,1}}$, which are purely thermal without the BEC component. The number fractions $N_{0,1}/(N_{-1}+N_0+N_1)$ increase with $t_h$, where $N_n$ is the atom number in $\ket{\xi_n}$; see Fig.~\ref{fig:thermal}a.
%
%
At $t_h=1.0$~s, the total thermal atom number fraction is about $22\%$. At a given $t_h$, $N_{0,1}$ decreases with increasing $|\delta|$ and rapidly decreases when $|\delta|/2\pi \gtrsim 1$~kHz~\cite{SChen2018a}; see Fig.~\ref{fig:thermal}b. Besides, $N_{-1}$ has an exponentially decaying lifetime of $2.7(1)$~s.

\section{Derivation of the angular momentum $L_z$ from the quadrupole mode precession} 
As a calibration for $L_z$, we measure the quadrupole mode precession angle $\theta$ of the atoms deloaded to $\ket{m_F=-1}$ with stable $L_z=0, -2\hbar$, respectively, for the hold time $0.1~\mu$s $<t_h<1.6$~s. This allows for converting the $\theta$ of a vortex-nucleated state to the $L_z$. To prepare the $\ket{m_F=-1}$ with $L_z=0, -2\hbar$, respectively, we load the condensate into the dressed state with canonical angular momentum $\ell=\ell_g=0$ and $-2\hbar$, respectively, hold $t_h$ and then deload the atoms to $m_F= -1$ with $L_z=\ell-\hbar=0$ and $-2\hbar$, respectively. (We prepare the $\ell_g=-2\hbar$ state by starting with atoms spin-polarized in $\ket{m_F=1}$ followed by adiabatic loading into the dressed state with $\delta/2\pi=-500$~Hz.) Since the initial $\ell$ equals to the absolute ground state's $\ell_g$, $\ell$ is unchanged during $t_h$, i.e., stable and without vortex nucleations. Therefore, the final $L_z$ remains $0,-2\hbar$ for $\ell_g=\hbar,-\hbar$, respectively, for all $t_h$.
The precession angle is
\begin{align}
	\theta=\frac{L_z}{2m R_{\bot}^2}(\tau+\tau_{\rm exp}),	
\end{align}
where $\theta_{\rm trap}=L_z/2m R_{\bot}^2 \tau$ is the precession angle in the trap for precession time $\tau$ given by a sum rule approach~\cite{SZambelli1998}. $R_{\bot}$ is the transverse size, $R_{\bot}^2= \langle x^2+y^2 \rangle$ and $\tau_{\rm exp}$ is an additional time accounting for the precession during TOF expansion. Here, the precession angle during the excitation of quadrupole mode is negligibly small. We perform 3D TDGPE simulations for the quandrupole mode precession with $0<\tau\leq 8.2$~ms and $23.9$~ms TOF for $N \approx 1.3 \times 10^5$ atoms and $L_z=0, \hbar, 2\hbar$, respectively. We find that $\theta_{\rm trap}\propto \tau$, $\theta_{\rm trap}\approx 14^{\circ}$ for ($L_z=\hbar,\tau=8.2$~ms) and $\tau_{\rm exp}$ contributes $\approx 5^{\circ}$ to the overall $\theta\approx 19^{\circ}$, and $\theta$ doubles for $L_z= 2\hbar$. Here, $\tau_{\rm exp}$ has significant contribution in the overall precession angle $\theta$.
%
%

\begin{figure}
	\centering
	\includegraphics[width=3.5 in]{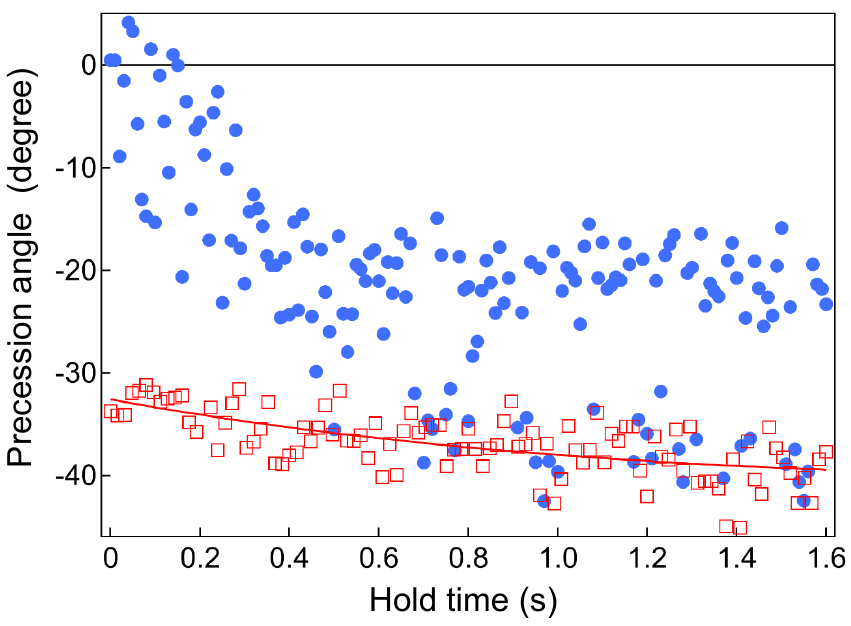}
	\caption{Experimental data of the precession angles of the surface quadrupole mode vs. hold time $t_h$. The blue dots are for detuning $\delta/2\pi=-500$~Hz, from which we obtained $L_z$ shown in Fig.~3 of the paper. The magnitude of the angle for calibration $\theta_c (-2\hbar)-\theta_c(0\hbar)$ depicted with red squares slightly increases with $t_h$ due to the decrease of BEC number and the size $R_{\bot}$.}
	\label{fig:Lz_calibration} 
\end{figure}

We calibrate $L_z$ from measured $\theta$ for atoms with definite $L_z=0, -2\hbar$, respectively, vs. hold time $t_h$. Then we apply linear interpolation to derive $L_z$ from $\theta$ without using the theoretical formula of $\theta_{\rm trap}$.
In our experimental data of calibration, $\theta_c(L_z=0,-2\hbar)$, we find two deviations from the simulations. First, $\theta(L_z=0)$ is theoretically zero, while our measured $\theta_c(L_z=0)$ is nonzero and slightly depends on $t_h$. Second, the angle difference $\theta_c(-2\hbar,t_h)-\theta_c(0\hbar,t_h)$ is about $85~\%$ of that in the simulation. We identify that the deviation is largely from the precession angle during TOF, instead of that in the trap. We believe this is attributed to imperfect laser beam alignment. We assume the factor for the deviation of $\theta_c(-2\hbar,t_h)-\theta_c(0\hbar,t_h)$ is independent of $L_z$, and derive $L_z$ from $\theta$ using linear conversions for all $t_h$,
\begin{align}
	L_z= -2\hbar \frac{\theta-\theta_c(0\hbar)}{\theta_c (-2\hbar)-\theta_c(0\hbar)},
\end{align}
see Fig.~\ref{fig:Lz_calibration}. The uncertainty of $L_z$ for the measurements of $L_z=0,-2\hbar$ within 15 shots, where $L_z$ is stable without vortex nucleations, is $\approx 0.2\hbar$.

\section{Cylindrical asymmetry of the Raman coupling}
In the ideal condition for the Raman beams, both LG and G beam are cylindrically symmetric and the Raman coupling is denoted as
\begin{align}
	\vec {\Omega}&=\Omega(r^{'})\cos\phi^{'}{\mathbf e}_{x}-\Omega(r^{'})\sin\phi^{'}{\mathbf e}_{y}+ \delta {\mathbf e}_{z},\\
	\Omega(r^{'})&=\Omega_M^{'}\sqrt{e}\frac{r^{'}}{r_M}e^{-{r^{\prime}}^2/2r_M^2},
\end{align}
where the cylindrical coordinate $(r^{'},\phi^{'})$ is with respect to the LG beam center $O^{'}$; the $(\cos\phi^{'},-\sin\phi^{'})$ is for the order-one LG beam phase winding. Ideally, $O^{\prime}$ is identical to the BEC center $O$. While practically, $O^{'}$ can slightly deviate from $O$: $O^{'}$ is displaced from $O$ by up to $0.4\micron$ with a random direction. Moreover, the intensity of the LG beam is not perfectly cylindrical symmetric with respect to $O^{'}$, and the actual Raman coupling is described by
\begin{align}
	\vec{\Omega}&=\Omega(r^{'},\phi^{'})\cos\phi^{'}{\mathbf e}_{x}-\Omega(r^{'},\phi^{'})\sin\phi^{'}{\mathbf e}_{y}+ \delta {\mathbf e}_{z},\\
	\Omega(r^{'},\phi^{'})&=\Omega_M^{'}\sqrt{e}\frac{r^{'}}{r_M}e^{-{r^{\prime}}^2/2r_M^2}\left[1+\sum_{\ell'=1}^{\ell'_{\rm max}} f(\ell') \cos(\ell'\phi^{'}+\eta_{\ell'})\right].
	\label{eq:Omega_shift}
\end{align}
The nonzero $f({\ell'})$ characterizes the asymmetry of the LG beam. We experimentally determined $\Omega_M^{'},f(\ell'),\eta_{\ell'}$ by measuring $\Omega(r^{'},\phi^{'})$ from BEC under Raman pulsing. We find that $f({\ell'}=1)=0.2$ and $f({\ell'})=0.1/\ell'$ for $2 \leq \ell' \leq \ell'_{\rm max}=15$, {$(\eta_1-\eta_2)/2\pi= 0.11 \pm 0.21$ and $\eta_{\ell'}$ is random within $[0,2\pi]$ for $3 \leq \ell' \leq \ell'_{\rm max}=15$.
	
	\section{Additional information of experimental data}
	\subsection{Experimental data labeled by effective detuning $\delta/\Omega(R_{\rm TF})$}
	
	In our main paper, we express the detuning $\delta/2\pi$ in unit of Hz. Actually, a physically relevant expression is the ratio between the detuning and the Raman coupling. Since Raman coupling $\Omega(r)$ depends on $r$, we choose to normalize $\delta$ to the Raman coupling at the periphery of the condensate, $\Omega(R_{\rm TF})$. The data with normalized detuning is shown in Fig.~\ref{fig:normalized_delta1} and Fig.~\ref{fig:normalized_delta2}. The detunings $\delta/2\pi=-200,-500,-600,-1000$~Hz in Fig.~3 correspond to $\delta/\Omega(R_{\rm TF})=-0.13,-0.33,-0.39,-0.66$, respectively.
	
	\begin{figure}
		\centering
		\includegraphics[width=4.0 in]{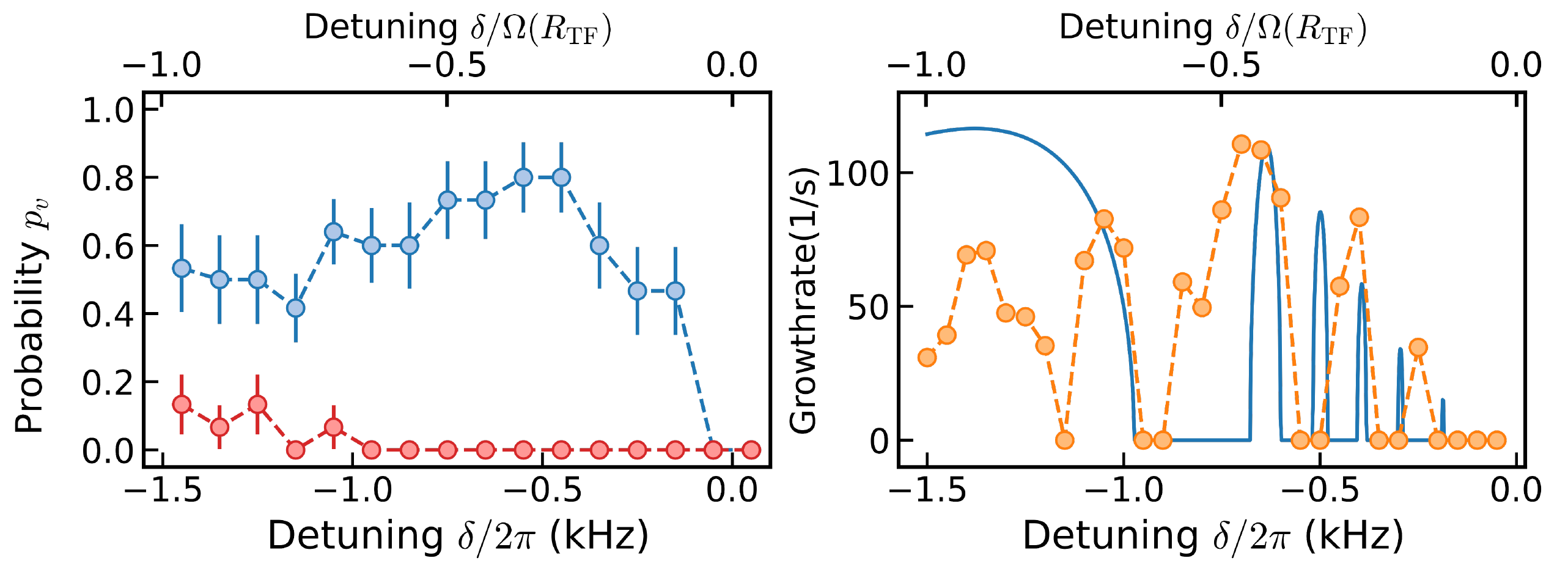}
		\caption{Data of Fig.~2a and Fig.~2c of the main paper with normalized detuning, $\delta/\Omega(R_{\rm TF})$.}
		\label{fig:normalized_delta1}
	\end{figure}
	
	\begin{figure}
		\centering
		\includegraphics[width=3.75 in]{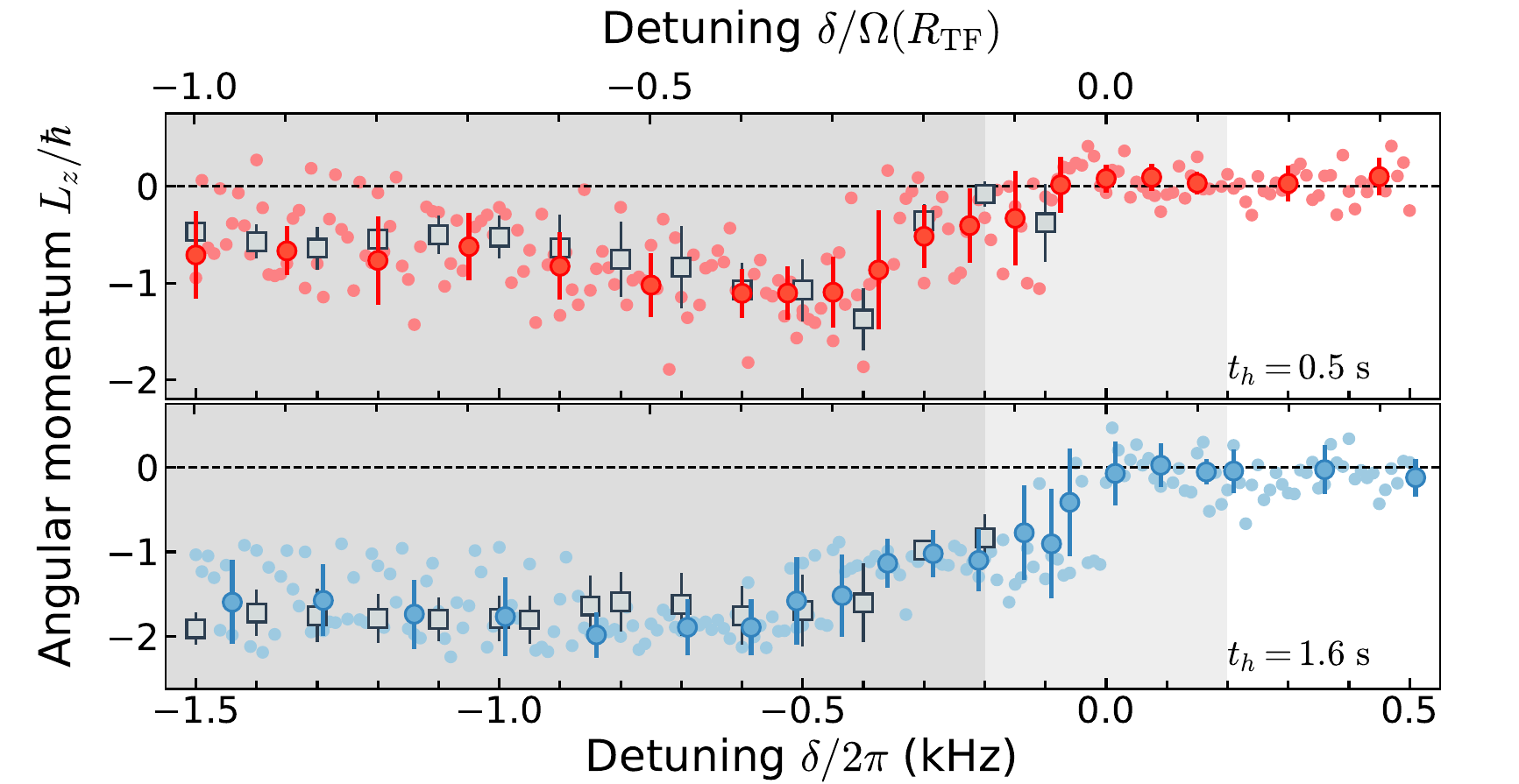}
		\caption{Data of Fig.~4a of the main paper with normalized detuning, $\delta/\Omega(R_{\rm TF})$.}
		\label{fig:normalized_delta2}
	\end{figure}

	\subsection{Vortex nucleation dynamics illustrated in energy dispersion}
	We prepare the initial condensate with canonical OAM $\ell=0$, therefore $\ell=0\neq \ell_g$ at detuning $\delta/2\pi<200$~Hz where vortex nucleations occur initiated by dynamical instability. The absolute ground state's cOAM $\ell_g$ depends on $\delta$, where $\ell_g=0,-\hbar,-2\hbar$ for $\delta/2\pi>200$~Hz, $|\delta/2\pi|<200$~Hz, and $\delta/2\pi<-200$~Hz, respectively (see Fig.~\ref{fig:groundstate}). Figure~\ref{fig:dispersion}a shows the energy dispersion, energy vs. $\ell$, for three example detunings: $\delta/2\pi= 400$~Hz with $\ell_g=0$, $\delta/2\pi= -100$~Hz with $\ell_g= -\hbar$, and $\delta/2\pi= -600$~Hz with $\ell_g= -2\hbar$. For a BEC with $\delta/2\pi= 400$~Hz, it is energetically stable, while in the case of $\delta/2\pi= -100$~Hz and $\delta/2\pi= -600$~Hz, $\ell$ evolves in time from $0$ to $-\hbar$ and from $0$ to $-2\hbar$ given sufficiently long hold time $t_h$, respectively. Figure~\ref{fig:dispersion}b shows the experimental data $L_z(t_h)$ of $\delta/2\pi= -600$~Hz (same as that in Fig.~3 of the main paper), where $L_z(t_h)=0$ and $L_z(t_h=1.6~{\rm s})$ reaches $\ell_g= -2\hbar$. Figure~\ref{fig:dispersion}c shows the experimental data $L_z$ vs. $\delta$ with $t_h=1.6$~s (same as that in Fig.~4 of the main paper). At $\delta/2\pi= -100$~Hz, $L_z(t_h=1.6~{\rm s})$ reaches $\ell_g= -\hbar$. Again we observe that $L_z(t_h=1.6~{\rm s})$ reaches $\ell_g= -2\hbar$ at $\delta/2\pi= -600$~Hz.

	\begin{figure}
		\centering
		\includegraphics[width=5.5 in]{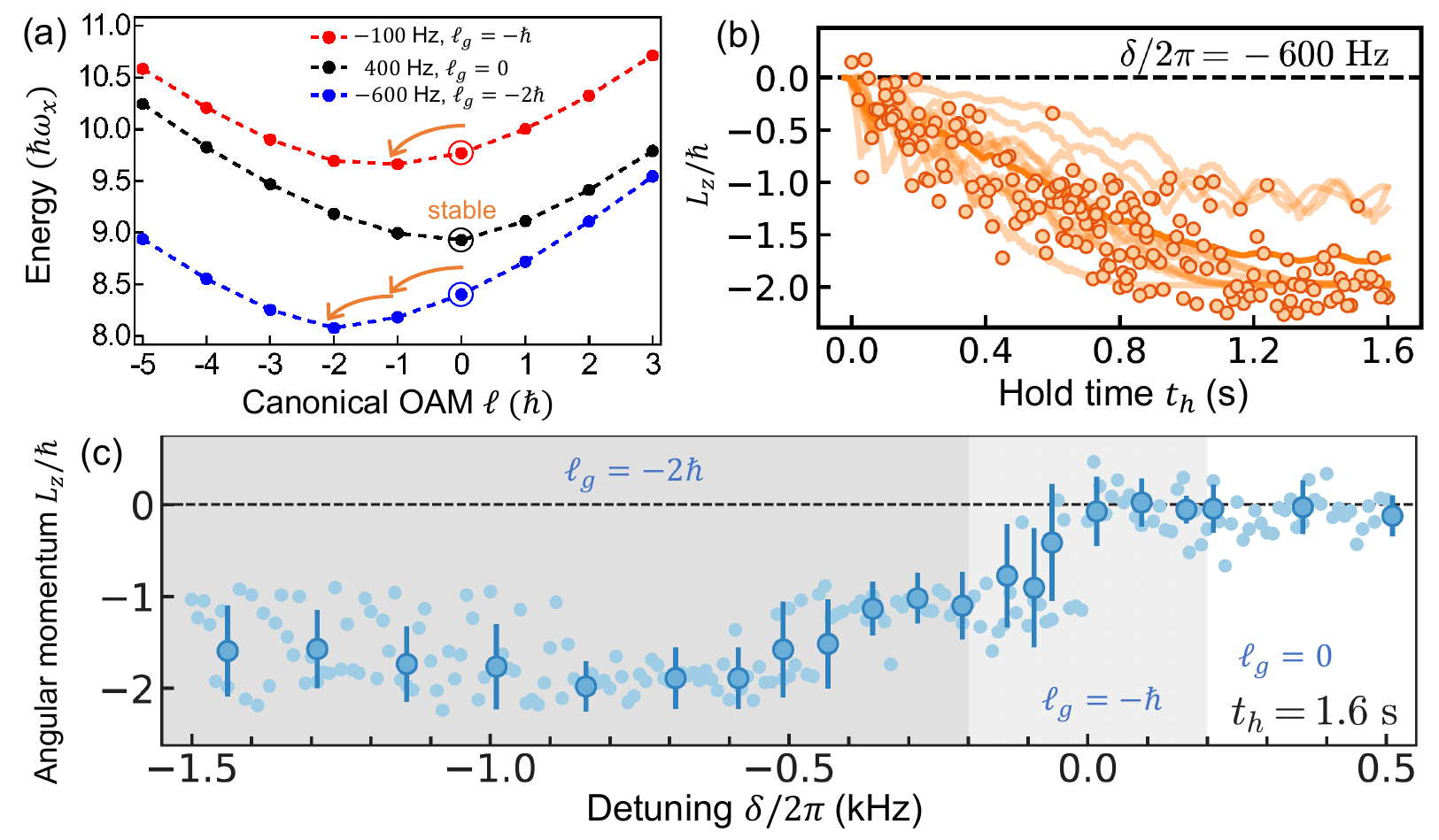}
		\caption{(a) Energy vs. cOAM $\ell$ for detuning $\delta/2\pi=400, -100, -600$~Hz with the ground state cOAM $\ell_g=0,-\hbar,-2\hbar$, respectively. For a BEC initially with $\ell=0$ (denoted by an additional outer circle), the state can dynamically evolve to $\ell_g=-\hbar,-2\hbar$ for $\delta/2\pi= -100, -600$~Hz, respectively. The BEC is stable at $\delta/2\pi=400$~Hz given that $\ell=\ell_g=0$. (b) experimental data $L_z$ vs. hold time $t_h$ (solid symbols) with $\delta/2\pi= -600$~Hz. $L_z$ can evolve to $\ell_g=-2\hbar$ at $t_h=1.6$~s. (c) experimental data $L_z$ vs. $\delta$ with $t_h=1.6$~s. At $\delta/2\pi= -100, -600$~Hz, $L_z(t_h=1.6~{\rm s})$ reaches $\ell_g= -\hbar, -2\hbar$, respectively.}
		\label{fig:dispersion}
	\end{figure}
	
	\section{Numerical simulations}
	\subsection{Bogoliubov-de Gennes Spectrum in 2D}
	We solve the Bogoliubov-de Gennes (BdG) equation in the two-dimensional (2D) system.
	We first numerically obtain a stationary solution with $\ell=0$ under the SOAMC, whose wave function is given by
	\begin{align}
		|\Psi^{(0)}(r,\phi)\rangle
		\equiv\begin{pmatrix}
			\psi^{(0)}_1(r,\phi)\\\psi^{(0)}_0(r,\phi)\\\psi^{(0)}_{-1}(r,\phi)
		\end{pmatrix}
		=\begin{pmatrix} G_1(r)e^{i2\phi} \\ G_0(r)e^{i\phi} \\ G_{-1}(r)\end{pmatrix}
		=\underbrace{\begin{pmatrix} e^{i2\phi} & 0 & 0 \\ 0 & e^{i\phi} & 0 \\ 0 & 0 & 1\end{pmatrix}}_{\equiv \mathcal{U}(\phi)}
		\begin{pmatrix} G_1(r) \\ G_0(r) \\ G_{-1}(r)\end{pmatrix}
		.
		\label{eq:OP0}
	\end{align}
	Here, $G_{0,\pm1}(r)$ are complex function of $r$, which are determined so that Eq.~\eqref{eq:OP0} satisfies the stationary Gross-Pitaevskii equation (GPE).
	We choose the interaction parameters for the 2D system such that the Thomas-Fermi radius along the radial direction coincides with that in 3D. The spinor part of the obtained $|\Psi^{(0)}(r,\phi)\rangle$ is slightly different from $|\xi_{-1}\rangle$ due to the kinetic energy term in Eq.~(S-1).
	
	We expand the order parameter around the obtained solution as
	\begin{align}
		|\Psi(r,\phi,t)\rangle=e^{i\mu t/\hbar}\mathcal{U}(\phi)\left\{\begin{pmatrix} G_1(r) \\ G_0(r) \\ G_{-1}(r)\end{pmatrix} + \sum_{q=0,\pm1,\pm2,\cdots}\bigg[e^{iq\phi-i\omega_qt}
		\underbrace{\begin{pmatrix} u_{1,q}(r) \\ u_{0,q}(r)\\ u_{-1,q}(r)\end{pmatrix}}_{\equiv|u_q(r)\rangle}
		+e^{-iq\phi+i\omega_q^*t}
		\underbrace{\begin{pmatrix} v^*_{1,q}(r) \\ v_{0,q}^*(r)\\ v^*_{-1,q}(r)\end{pmatrix}}_{\equiv|v_q^*(r)\rangle}
		\bigg]\right\}
		\label{eq:OP}
	\end{align}
	By substituting Eq.~\eqref{eq:OP} into the time-dependent GPE (TDGPE), and neglecting higher-order terms with respect to $u_{m,q}(r)$ and $v_{m,q}(r)$, we obtain the BdG equation, which is written as in the following form:
	\begin{align}
		\mathcal{H}_q\begin{pmatrix}
			|u_{q}(r)\rangle \\ |v_{q}(r)\rangle
		\end{pmatrix}
		&=\hbar\omega_{q}
		\begin{pmatrix}
			|u_{q}(r)\rangle \\ |v_{q}(r)\rangle
		\end{pmatrix},
		\label{eq:BdG_q}\\
		\mathcal{H}_q&\equiv
		\begin{pmatrix}
			\textrm{H}_q & \textrm{H}_{\rm od}\\ -\textrm{H}_{\rm od}^* & -\textrm{H}_{-q}
		\end{pmatrix},
	\end{align}
	where
	$\textrm{H}_q$ and $\textrm{H}_{\rm od}$ are $3\times 3$ Hermitian and symmetric matrices, respectively.
	Note that because we consider a circularly symmetric system, the BdG equation is block diagonal for each $q$.
	In addition, because of the particle-hole symmetry, 
	i.e., $\mathcal{C}\mathcal{H}_q\mathcal{C}^{-1}=-\mathcal{H}_{-q}$
	where $\mathcal{C}=\tau_xK$ with $\tau_x$ being the Pauli matrix in the Nambu space and $K$ the complex conjugate operator,
	a state obtained by applying to $\mathcal{C}$ to
	an eigenstate of $\mathcal{H}_q$ with eigenvalue $\hbar\omega_q$ 
	is an eigenstate of $\mathcal{H}_{-q}$ with eigenvalue $-\hbar\omega_{-q}$. It follows that the eigenmodes with $q$ and $-q$ are obtained from a single eigenvalue equation, and hence they are coupled.
	
	We numerically solve the BdG equation for each $q$ and obtain $\omega_{q,n}$ where $n$ is the energy level index.
	Figure~\ref{fig:BdG_q} shows the BdG spectrum for $-2\le q\le 2$. The lowest eigenfrequencies of $q=-2$ and $-1$, $\omega_{-2,0}$ and $\omega_{-1,0}$, become negative in a certain range of $\delta$, and we plot $-\omega_{-2,0}$ and $-\omega_{-1,0}$ together with $\omega_{2,n}$ and $\omega_{1,n}$, respectively. 
	They have nonzero imaginary part when ${\rm Re}[\omega_{q,n}+\omega_{-q,0}]\sim 0$~\cite{Kawaguchi2004}.

	\begin{figure}
		\centering
		\includegraphics[width=0.8\linewidth]{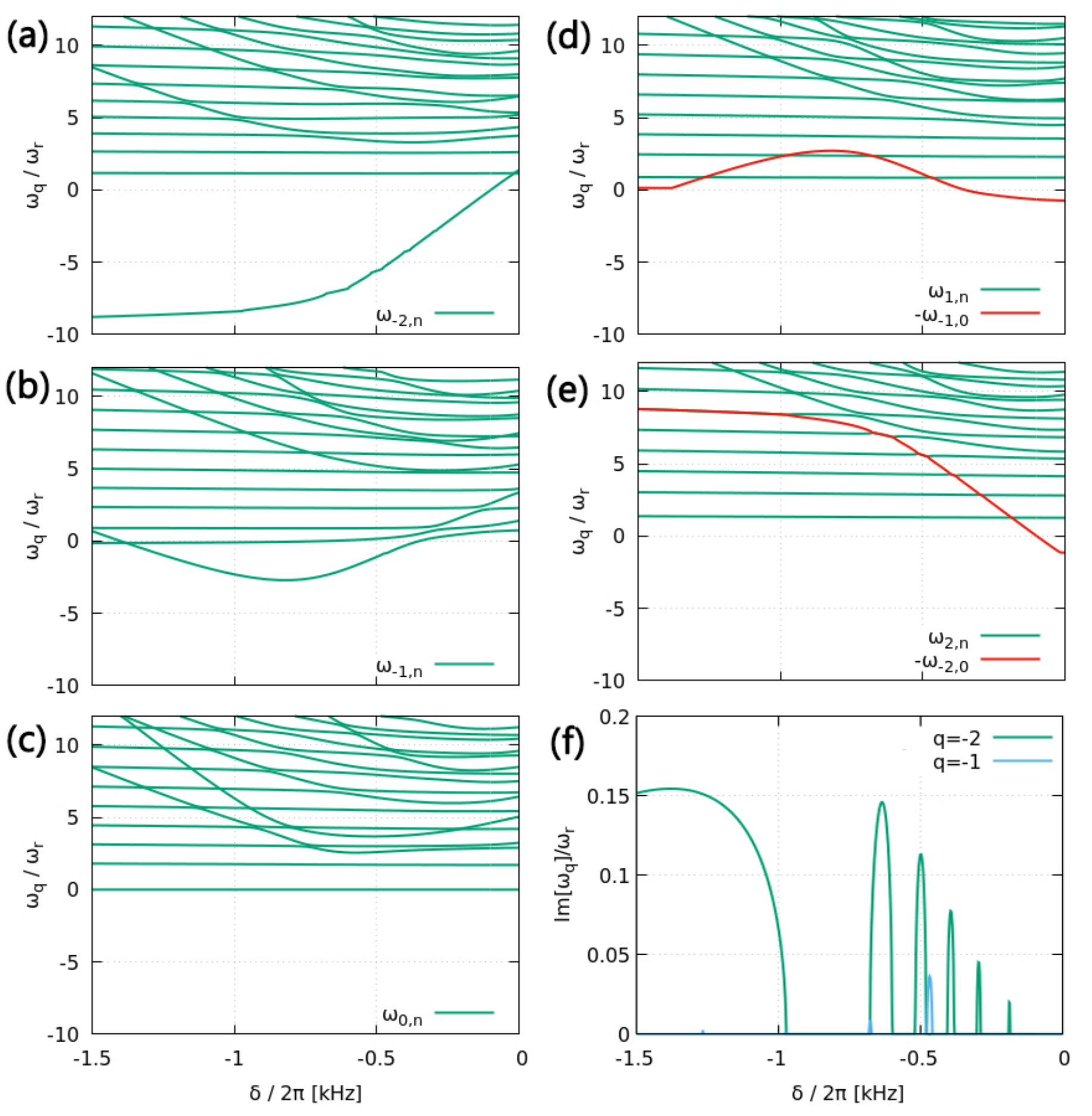}
		\caption{(a)-(e) Real parts of the BdG spectrum for $q=-2$ (a), $-1$ (b), $0$ (c), $1$ (d), and $2$ (e). (f) Imaginary parts of the eigenfrequency. In panels (d) and (e), $-\omega_{-1,0}$ and $-\omega_{-2,0}$ are also shown, respectively. Imaginary part arises when ${\rm Re}[\omega_{q,n}+\omega_{-q,0}]=0$.}
		\label{fig:BdG_q}
	\end{figure}
	
	The imaginary part mainly comes from the coupling between the $q=2$ and $-2$ modes. Here, $\omega_{-2,0}$ becomes largely negative as $\delta$ decreases. 
	This is a localized mode at the center of the condensate:
	Since the atoms just after the loading process at largely negative $\delta$ are almost in the $|m_F=+1\rangle$ component, which has the phase winding $e^{i2\phi}$ [see Eq.~\eqref{eq:OP0}], the condensate density is strongly suppressed at $r\sim 0$; The lowest-eigenfrequency mode of $q=-2$ is the localized mode at the density dip.
	Hence, the instability associated with this mode affects mainly the center of the condensate.
	In Fig.~\ref{fig:BdG_snapshots}, we show the wave function $|\Psi^{(0)}(r,\phi)\rangle$ at $\delta/2\pi=-450$ Hz (a) and
	\begin{align}
		|\Psi(r,\phi)\rangle
		=|\Psi^{(0)}(r,\phi)\rangle
		+0.05\,
		\mathcal{U}(\phi)\left[e^{-i2\phi}|u_{-2,0}(r)\rangle+e^{i2\phi}|v^*_{-2,0}(r)\rangle\right],
		\label{eq:psi0+BdG}
	\end{align}
	at $\delta/2\pi=-450$ (b) and $-500$ Hz (c), for which $\omega_{-2,0}$ has zero and nonzero imaginary part, respectively.
	Here, top and bottom panels show the density and phase profiles of the projected wave function onto $|\xi_{-1}\rangle$, which is defined by Eq.~\eqref{eqn:xi_minus1}.
	One can see from Fig.~\ref{fig:BdG_snapshots}(b) that the core mode with $q=-2$ creates two density dips around the trap center, which correspond to two vortices with phase winding $+1$. In addition, due to our gauge choice, an additional phase winding of $-2$ appears at the trap center, resulting in the appearance of two vortex-antivortex pairs after deloading.
	When ${\rm Im}\,\omega_{-2,0}\neq 0$ [Fig.~\ref{fig:BdG_snapshots}(c)], the density at the medium radius is also modulated due to the coupling with the $\omega_{2,n}$ mode.
	Although the unstable mode modulates the condensate density in the middle radius, it is hard to observe in experiments.
	
	\begin{figure}
		\centering
		\includegraphics[width=3.5 in]{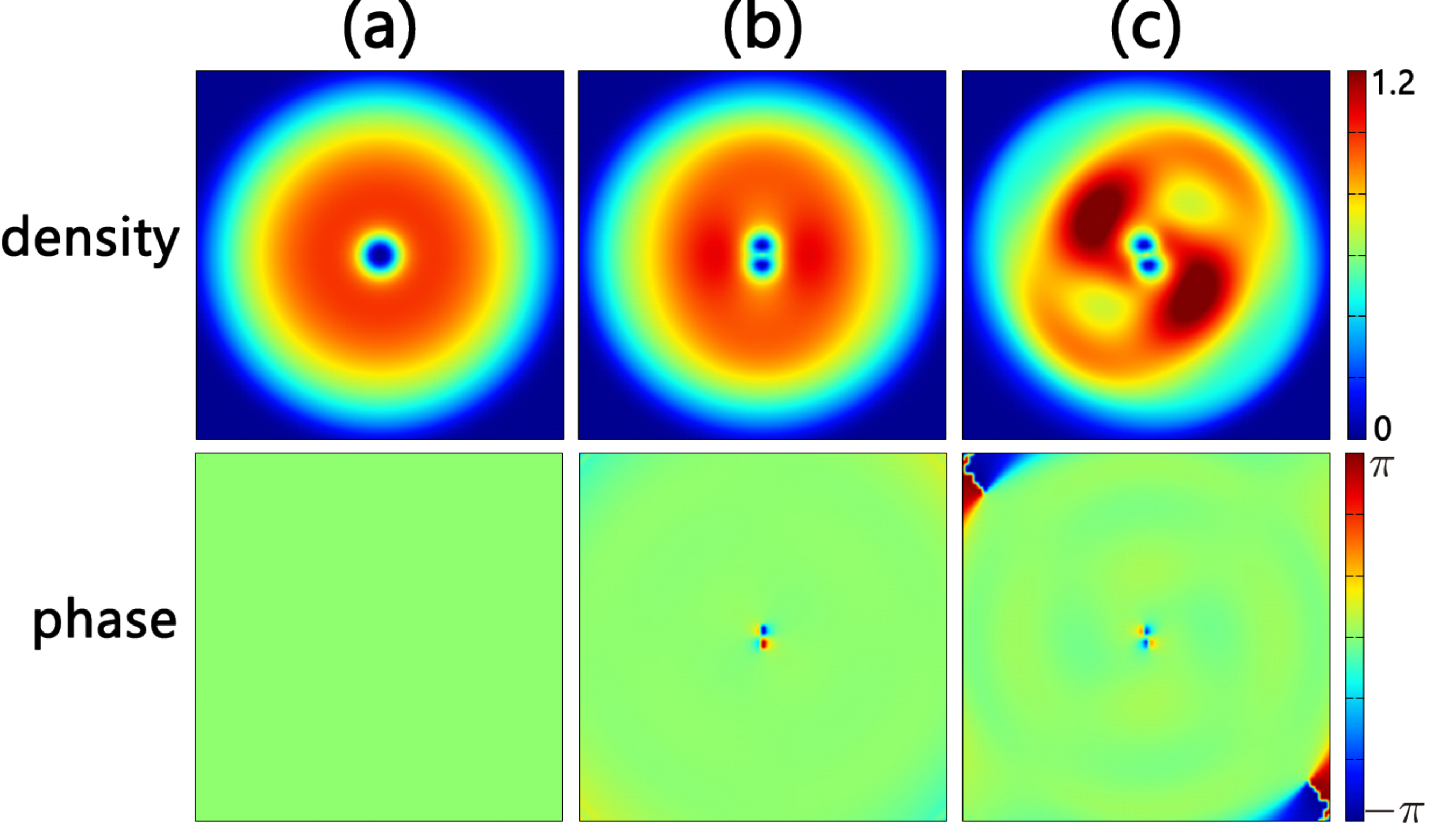}
		\caption{Wave function of the stationary state with $\ell=0$, $|\Psi^{(0)}(r,\phi)\rangle$, at $\delta/2\pi=-450$ Hz and those with the lowest-eigenfrequency BdG modes with $q=-2$, Eq.~\eqref{eq:psi0+BdG}, at $\delta/2\pi=-450$ (b) and $-500$ Hz (c), at which $\omega_{-2,0}$ has zero and nonzero imaginary part, respectively. Shown are the density (top panels) and phase (bottom panels) profiles of the projected wave function onto the lowest-energy dressed state $|\xi_{-1}\rangle$. The density is normalized by the maximum of the total density without the BdG mode.
		}
		\label{fig:BdG_snapshots}
	\end{figure}

	\subsection{Dynamical instabilities in 3D}
	In order to confirm the appearance of dynamical instability in 3D, we numerically simulate the TDGPE and investigate the growth of fluctuations.
	We first calculate a stationary state under the SOAMC with a given $\delta$ by fixing $\ell$ to be zero, where the wave function for $|m_F\rangle$ component is written as $g_{m_F}(r,z)e^{i(m_F+1)\phi}$.
	Here, we use the same harmonic potential as that in the experiment.
	We then prepare the initial order parameter by adding an $q=-2$ component as
	\begin{align}
		|\Psi^\textrm{(3D,ini)}(x,y,z)\rangle=\left[\begin{pmatrix}
			g_1(r,z)e^{2i\phi} \\ g_0(r,z)e^{i\phi} \\ g_{-1}(r,z)
		\end{pmatrix}
		+ \Delta_\textrm{noise} \,n(0,z)e^{-r^2/\xi^2}\begin{pmatrix} 1\\ 0 \\ 0\end{pmatrix}\right],
		\label{eq:3Dini}
	\end{align}
	where $n(0,z)=|g_{-1}(r=0,z)|^2$ is the number density at $r=0$, and we use $\xi=2.4~\mu$m and $\Delta_\textrm{noise}=0.05$ ( $\Delta_\textrm{noise}=0.2$) for $\delta/2\pi>-0.8$~kHz ($\delta/2\pi\le-0.8$~kHz).
	Starting from the above initial state, we calculate the time evolution of the condensate $|\Psi(x,y,z,t)\rangle$ 
	with monitoring the angular Fourier component of the projected wave function onto $|\xi_{-1}\rangle$,
	\begin{align}
		\varphi_{q}=\int e^{-iq\phi}\langle \xi_{-1}|\Psi(x,y,z,t)\rangle dxdydz.
		\label{eq:varphi_q}
	\end{align}
	
	Being consistent with the 2D BdG results, $|\varphi_{-2}|^2$ exhibits exponential growth for certain values of $\delta$.
	Figure~\ref{fig:varphi-2} shows the examples of the time evolution of $|\varphi_{-2}|^2$. 
	We fit $\log|\varphi_{-2}|^2$ for each $\delta$ with a function $f(t)=a+2bt$ in a certain region of $t$.
	The 3D TDGPE data in Fig. 2c of the main paper shows such obtained $b$ for each $\delta$.
	
	\begin{figure}
		\centering
		\includegraphics[width=3.8 in]{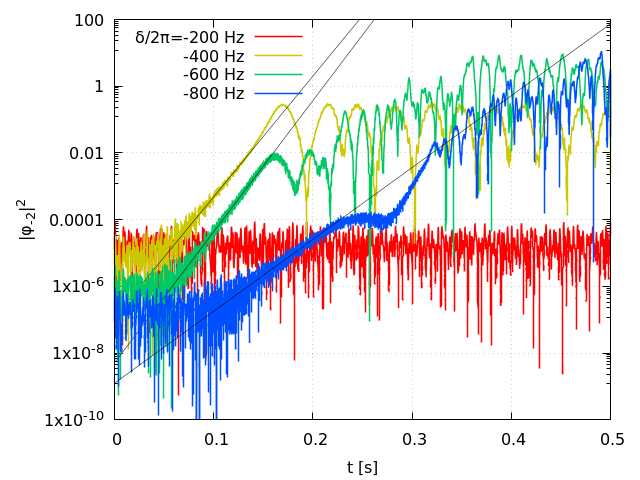}
		\caption{Time evolution of the $q=-2$ angular Fourier component of the wave function, Eq.~\eqref{eq:varphi_q}, starting from the initial wave function given in Eq.~\eqref{eq:3Dini}.
			The thin black lines depict the fitting functions $e^{a+bt}$ at $\delta/2\pi=-400, -600$, and $-800$~Hz.
		}
		\label{fig:varphi-2}
	\end{figure}
	
	Note that the above analysis is possible only for a circularly symmetric system. With cylindrical asymmetries, $q$ is no longer a good quantum number that characterizes the BdG mode, and we cannot expect the exponential growth of the angular Fourier component. Because symmetric states are generally more stable, we expect the region with dynamically unstable modes to expand in the presence of cylindrical asymmetry.

	\subsection{Numerical simulations for long-time dynamics}
	\label{sec:simulation_Fig3}
	In the simulation for Fig.~3, we first calculate the ground-state wave function, $\psi^{(0)}(\vec{r})$, of a BEC in the $|m_F=1\rangle$ state before loading the SOAMC by the imaginary-time propagation method. We then add a random noise in each spin component and prepare the initial state as
	\begin{align}
		|\Psi(\vec{r},t=0)\rangle = \psi^{(0)}(\vec{r})\left[\begin{pmatrix} 0 \\ 0 \\ 1 \end{pmatrix}+ \begin{pmatrix}\alpha_{1,r}+i\alpha_{1,i} \\ \alpha_{0,r}+i\alpha_{0,i} \\ \alpha_{-1,r}+i\alpha_{-1,i}\end{pmatrix}\right],
	\end{align}
	where $\alpha_{m_F,i/r}$ are uniform random numbers between $-0.05$ and $0.05$ independently determined on each grid of size $dx=dy=0.17~\mu$m and $dz=0.70~\mu$m. Starting from the above wave function, we load the SOAMC and investigate the evolution of the condensate during the holding time.

	Although dynamical instability triggers the instability, it cannot largely change the total OAM as observed in the experiment. Within the BdG analysis, because the eigenmode with a complex eigenfrequency satisfies $\int 2\pi rdr \left[\langle u_{q,n}(r)|u_{q,n}(r)\rangle-\langle v_{q,n}(r)|v_{q,n}(r)\rangle\right]=0$, the growth of the unstable mode does not lead to the change in $L_z$. In addition, a linearly unstable system does not always have nonlinear instability.
	In the present case, even when we see the appearance of two density dips in the projected wave function, they remain close at the center of the condensate.
	
	After some calculations, we find that energy dissipation and cylindrical asymmetry are needed to reproduce the experimental results.
	The energy dissipation is phenomenologically introduced by replacing $i\partial/\partial t$ in the TDGPE
	with $(i-\gamma)\partial/\partial t$ with manually keeping the total number of atoms constant.
	For the simulations in Fig.~3 of the main paper, we choose $\gamma=0.003$.
	As for asymmetry, we incorporate two asymmetries that originally existed in the experiment:
	One is the shift of the LG beam from the trap center, and the other is the asymmetric power profile of the LG beam, which are included in the $x,y$ dependence on the effective magnetic field as
	\begin{align}
		\vec{\Omega}_{\rm eff}(x,y)&=\Omega(r',\phi')(\cos\phi',-\sin\phi',-\delta),\\
		r'&=\sqrt{[x-X(t)]^2+y^2}, \\
		\phi'&={\rm arg}[x-X(t)+iy],
		\label{eq:lfluc}
	\end{align}
	where $\Omega(r',\phi')$ is given in Eq.~\eqref{eq:Omega_shift}. Here, without loss of generality, we choose the direction of the LG beam shift along the $x$ axis and describe the amount of shift as $X(t)=X_0 \cos(2\pi \nu_0 t + \eta_0)$ with $X_0=0.4~\mu$m, $\nu=0.1$~Hz, and $\eta_0$ being a uniform random number between 0 and $2\pi$. We also use uniform random numbers between 0 and $2\pi$ for $\eta_{\ell'}$ of $1\le \ell'\le 15$ in Eq. (S-12) except for $\ell'=2$, for which we use a Gaussian random number with average $\eta_1-0.11\times 2\pi$ and standard deviation $0.21\times 2\pi$. In Fig.~3, the different results for 10 individual simulations for each detuning are due to two factors: First, the initial random noise in the order parameter; Second, the asymmetries of the LG beam due to its center position shift and noncircular symmetric intensity profile with random phases $\eta_0$ and $\eta_{\ell'}$, respectively.

	\begin{figure}
		\centering
		\includegraphics[width=6.5 in]{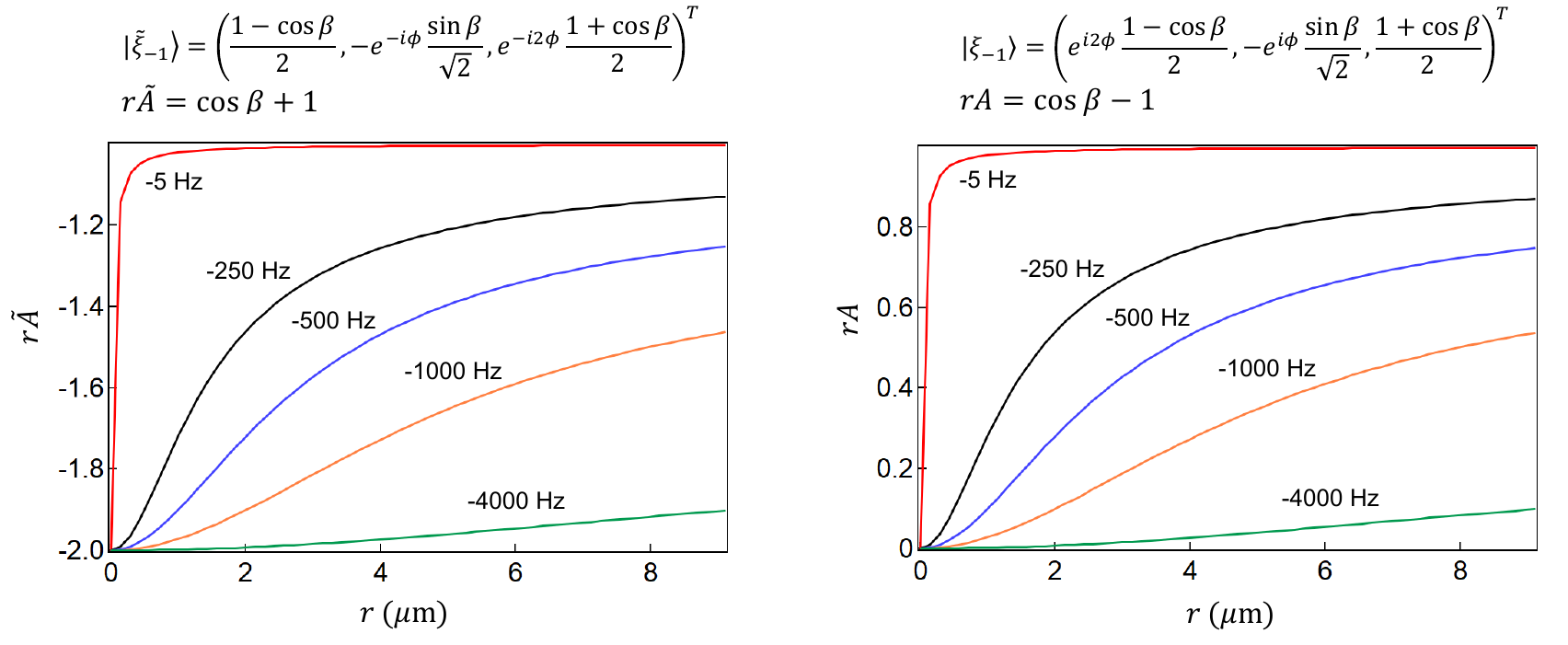}
		\caption{$r\tilde{A}$ (left panel) and $rA$ (right panel) vs. radial position $r$ with negative detunings, $\delta/2\pi=-5,-250,-500,-1000,-4000$ Hz, where the azimuthal gauge potential $\tilde{A}$ and $A$ have the phase gauge of $\ket{\tilde{\xi}_{-1}}$ and $\ket{\xi_{-1}}$, respectively.}
		\label{fig:two_gauges}
	\end{figure}
	
	\subsection{Vortex splitting dynamics in the spinor basis under gauge transformation}
	Our paper presents vortex nucleations by observing the atomic wave function $\varphi$ in the basis of lowest-energy dressed eigenstate $\ket{\xi_{-1}}$. It reveals that vortices nucleate from the cloud center of a vortex-free state with canonical momentum $\vec{p} = 0$. On the other hand, when we focus on the bare spin $\ket{m_F=1}$ component, a doubly quantized vortex is imprinted at $t_h=0$, which is dynamically unstable against splitting in the limit of $\delta\to-\infty$~\cite{SPu1999,SMottonen2003}. In our system, a similar instability exists even at a small negative $\delta$, and thus we can describe the dynamics in our experiment as vortex splitting dynamics by choosing a gauge differing from $\ket{\xi_{-1}}$ by a factor of $e^{-i2\phi}$, 
	\begin{align}
		\ket{\tilde{\xi}_{-1}} &= \left( \frac{1 -
			\cos \beta}{2}, -e^{-i \phi}\frac{\sin \beta}{\sqrt{2}},e^{-i 2\phi}\frac{1 +
			\cos \beta}{2} \right)^{\rm T} {\rm with~} \bar{\theta}+\bar{\gamma}=-\phi.
		\label{eqn:xi_tilde}
	\end{align}
	With the basis $\ket{\tilde{\xi}_{-1}}$, the spinor order parameter is described as $\tilde{\varphi}(\vec{r})\ket{\tilde{\xi}_{-1}}$, and the
	initial wave function $\tilde{\varphi}$ just after loading the dressed state has a doubly quantized vortex under the synthetic gauge potential $i\hbar\langle\tilde{\xi}_{-1}|\vec{\nabla}|\tilde{\xi}_{-1}\rangle= \tilde{A}\hat{\phi}=\hat{\phi}\hbar(\cos \beta+1)/r$; see Fig.~\ref{fig:two_gauges}a. When the detuning $\delta$ is a large negative value, $|r\tilde{A}|$ almost vanishes in the whole region of the cloud (see the curve with $\delta/2\pi=-4000$~Hz in Fig.~\ref{fig:two_gauges}a) where a doubly quantized vortex is unstable. On the other hand, when $\delta$ is negative and close to $0$, $|r\tilde{A}|$ vanishes at $r \sim0$ but approaches to a substantial value $\leq \hbar$ at the periphery of the condensate; Here, although the doubly quantized vortex is unstable at $r \sim 0$, the gauge potential $ \tilde{A}$ in the whole BEC substantially differs from zero. Therefore, when the doubly quantized vortex splits, the split vortices are difficult to leave the condensate because a phase winding $+1$ vortex in $\tilde{\varphi}$ is energetically metastable. See next subsection ``Metastable $L_z=-\hbar$ state'' for computation of the energetic stability. Such tendency for small negative $\delta$ is shown in the data of $L_z$ vs. $\delta$ with $t_h=1.6$~s in Fig.~4a: for $\delta/2\pi \gtrsim -500$~Hz, the system can not reach the ground state with $L_z=-2\hbar$, which is equivalent to a vortex-free $\tilde{\varphi}=e^{i2\phi}\varphi \propto 1$ indicating two split vortices leave the BEC; the state instead is trapped in a metastable $L_z=-\hbar$, which is equivalent to $\tilde{\varphi}=e^{i2\phi}\varphi \propto e^{i\phi}$ carrying a $+1$ vortex indicating one vortex has not left the BEC.
	
	\begin{figure}
		\centering
		\includegraphics[width=6 in]{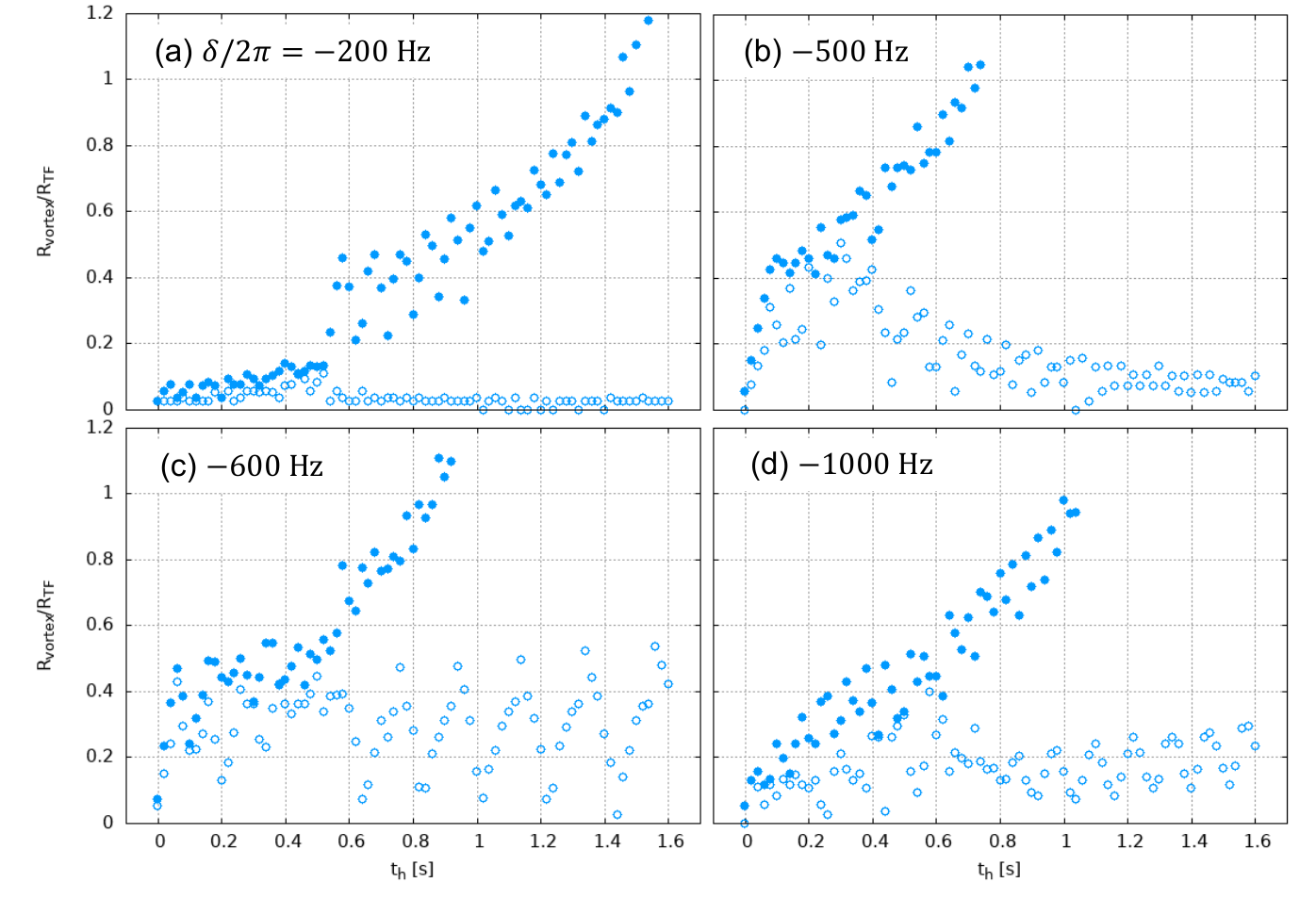}
		\caption{Vortex positions vs. hold time in the wave function $\tilde{\varphi}$, which is the spinor order parameter projected to $\ket{\tilde{\xi}_{-1}}$. Data shown are from one of the 10 simulations displayed in Fig.~3 of the main text. Filled (open) circles indicate the distance of the outer (inner) vortex from the trap center. The vertical axis is scaled by the Thomas-Fermi radius, $R_{\rm TF}$, in the radial direction.}
		\label{fig:vortexposition}
	\end{figure}
	
	In Fig.~\ref{fig:vortexposition}, we show the typical time evolution of the vortex positions in $\tilde{\varphi}$. Here, we obtain $\tilde{\varphi}$ by projecting the time-dependent 3D 3-component wave function onto $|\tilde{\xi}_{-1}\rangle$ and find the vortex positions on the $z=0$ plane. Shown are the result for the one of the 10 simulations shown in Fig.~3 of the main text, where the filled (open) circles indicate the distance of the outer (inner) vortex from the trap center. At a small $t_h$, the two vortices are at a similar distance from the center, and then the outer vortex goes out from the condensate. On the other hand, the inner vortex remains at the trap center [(a) $\delta/2\pi=-200$ Hz and (b) $-500$ Hz], oscillates around the trap center [(c) $\delta/2\pi=-600$ Hz], or slowly goes out from the condensate [(d) $\delta/2\pi=-1000$ Hz]. The different behavior of the inner vortex comes from the spatial dependence of the gauge field $\vec{A}(\vec{r})$.
	
	\begin{figure}
		\centering
		\includegraphics[width=3.5 in]{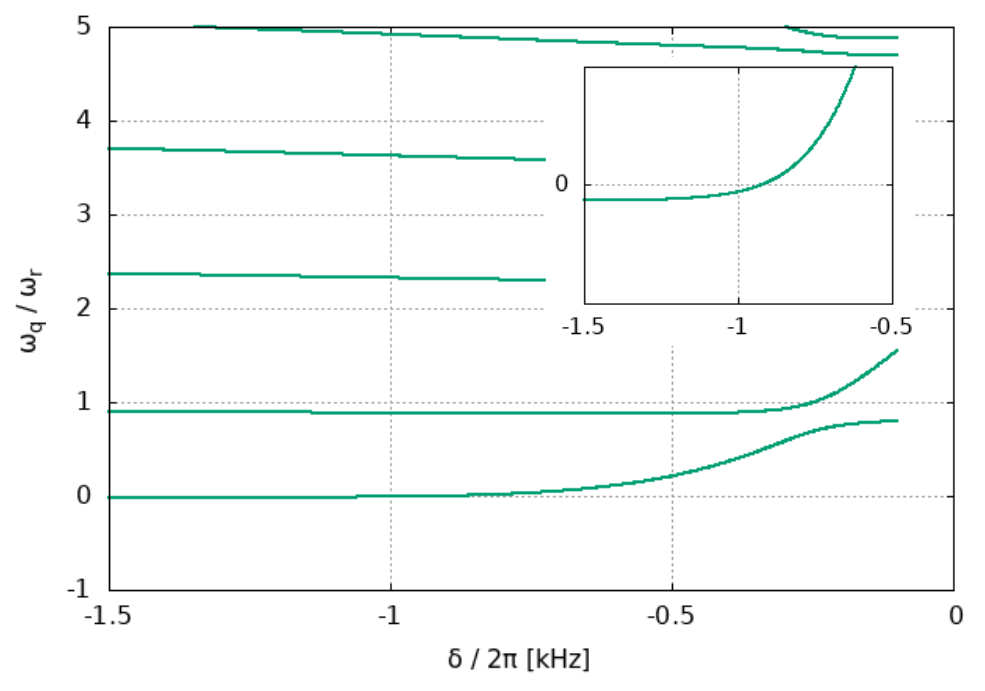}
		\caption{BdG eigenenergy spectrum $\omega_q$ in 2D, where the condensate mode is the dressed state with cOAM $\ell=-\hbar$. The lowest-energy BdG mode $\omega_0$ is positive at detuning above $-900$~Hz, which has Landau stability and corresponds to an energy barrier in the order parameter space. The inset is the magnified view of the main panel around $\omega=0$.}
		\label{fig:BdG_elln1}
	\end{figure}
	
	\begin{figure}
		\centering
		\includegraphics[width=3.5 in]{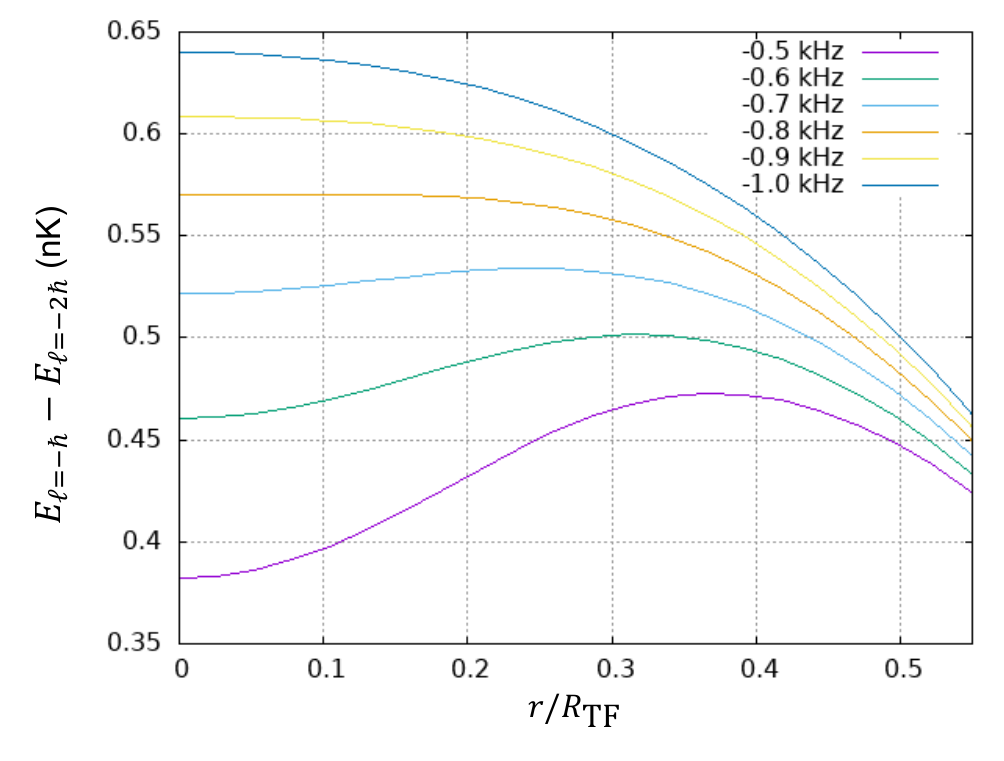}
		\caption{Energy landscape of the total energy of the metastable $\ell=L_z= -\hbar$ state in 2D relative to that of the ground state $\ell_g= -2\hbar$ vs. vortex radial position with detunings $\delta/2\pi=-0.5,-0.6,-0.7,-0.8,-0.9,-1.0$~kHz. There is a metastable energy minimum at $r=0$ for detuning $\delta/2\pi \gtrsim -800$~Hz.}
		\label{fig:landau_stability}
	\end{figure}
	
	\subsection{Metastable $L_z=-\hbar$ state}
	In the previous subsection, we describe that for a small negative detuning $\delta$, it is difficult for both the split vortices from an initially doubly quantized vortex to leave the BEC observed in $\ket{\tilde{\xi}_{-1}}$ basis. This corresponds to a metastable state where $\tilde{\varphi}$ carries one winding $+1$ vortex, equivalent to $\varphi$ with cOAM $L_z=-\hbar$. To explain the metastable $L_z=-\hbar$ state, we compute the 2D energetic (Landau) stability of a condensate with $L_z=-\hbar$ and find the BdG eigenenergy is positive at detuning above $-900$~Hz (see Fig.~\ref{fig:BdG_elln1}),  which suggests the existence of an energy barrier. Note that this 2D calculation employs a line density which gives the same radial TF radius as that in 3D. We also compute the energy landscape of the total energy of the metastable $\ell=L_z= -\hbar$ state in 2D relative to that of the ground state $\ell_g= -2\hbar$ vs. vortex radial position as displayed in Fig.~\ref{fig:landau_stability}. It shows a metastable energy minimum at $r=0$ for detuning $\delta/2\pi \gtrsim -800$~Hz, agreeing with the BdG spectrum. As the detuning becomes more negative, the energy dip at $r=0$ becomes shallower and narrower (dip within a smaller range of $r$), suggesting that a vortex can escape from the metastable state more easily if it is off-centered, even though there is Landau stability: In the $m_F=1$ component, during the dynamics of the other winding $+1$ vortex going out of the condensate, the remaining winding $+1$ vortex also moves and shifts from the condensate center, which corresponds to a $-1$ vortex shifted from the center with $L_z\approx -\hbar$ state after deloading. Because the Landau instability and the energy landscape depend on the line density used in the 2D calculation, we may not quantitatively compare the results of such a 2D calculation to our experimental data. However, we may conclude that the metastable $L_z= -\hbar$ state is more robust against the fluctuation of vortex position for a larger (less negative) detuning, which is $-400$~Hz $\lesssim \delta/2\pi \lesssim -200$~Hz in our data shown in Fig.~4a with $t_h=1.6$~s.

	\subsection{Initial deformation at the onset of vortex nucleations}
	In Fig.~\ref{fig:initial_deform_simu}, we show the detailed information of the wave function shown in Fig.~5 in the main paper: We plot the individual density and phase profiles of the wave function of all $m_F=-1,0, 1$ components, the ones for the projected wave function to $|\xi_{-1}\rangle$, and the ones in $|m_F=-1\rangle$ after deloading.

	\begin{figure}
		\centering
		\includegraphics[width=6.0 in]{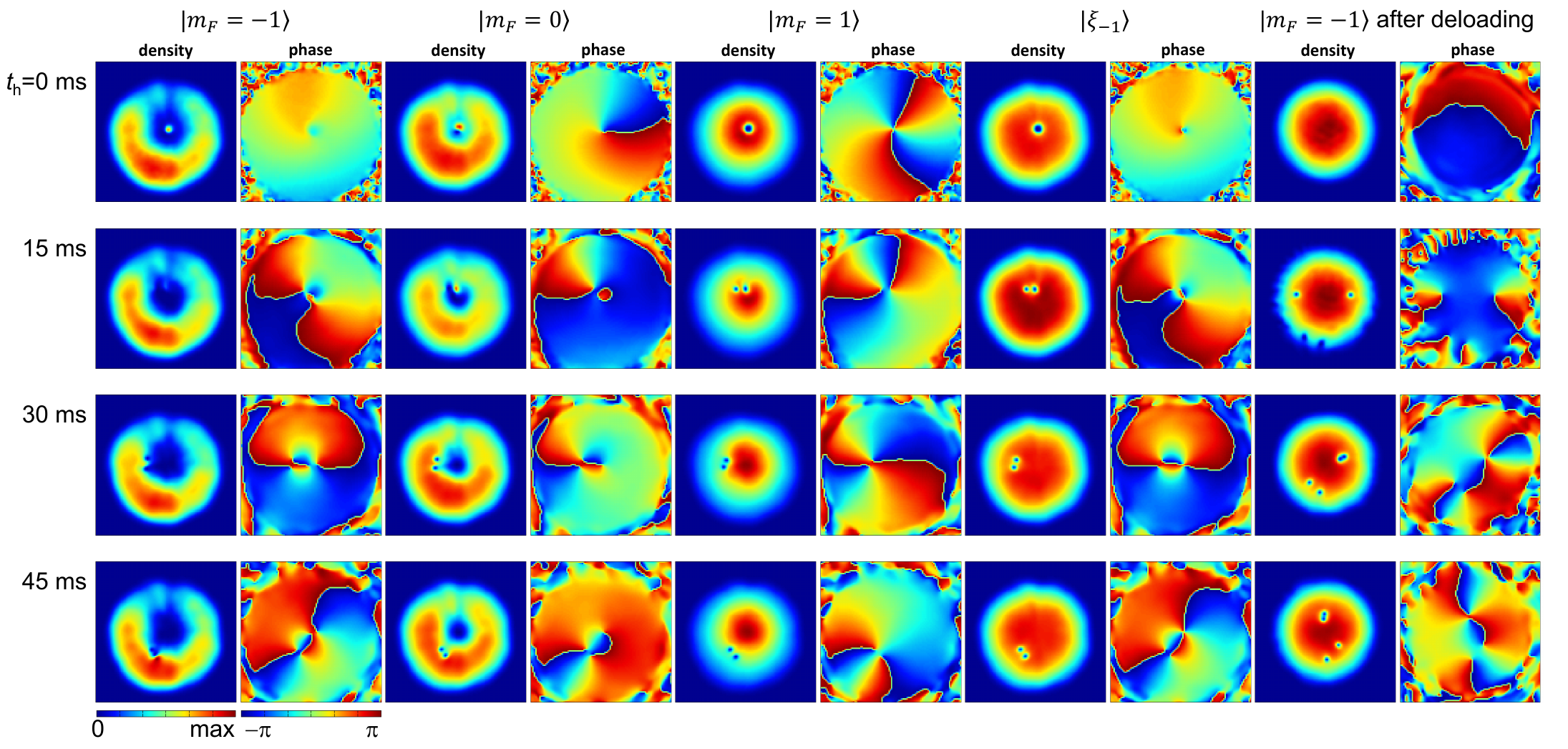}
		\caption{{\it in-situ} time evolution during the holding time at $\delta/2\pi=-600$ Hz numerically calculated with 3D TDGPE including asymmetry and energy dissipation. Shown are density $|\psi_\alpha(x,y,0,t)|^2$ and phase ${\rm arg}[\psi_\alpha(x,y,0,t)]$ profiles in the $z=0$ plane, where $\psi_\alpha(x,y,z,t)\equiv\langle \alpha|\Psi(x,y,z,t)\rangle$ is the projection of the 3-compoent spinor order parameter $|\Psi(x,y,z,t)\rangle$ onto the bare spin $m_F$ state ($|\alpha\rangle=|m_F=-1,0,1\rangle$) and the lowest-energy dressed state ($|\alpha\rangle=|\xi_{-1}\rangle$). The color scale for the density profile is given by the maximum value in each panel. The panel size is $17.4~\mu\textrm{m}\times17.4~\mu\textrm{m}$.
		}
		\label{fig:initial_deform_simu}
	\end{figure}
	
	When we deload the dressed state to a positive final detuning $\delta_{\rm del}$, 
	the additional phase winding becomes physical, i.e., a vortex with winding number $-2$ is imprinted in the $m_F=-1$ component. 
	In Fig.~\ref{fig:deloading_asym_simu}, we show the order-parameter change during the deloading process starting from $t_h=15$ ms in Fig.~\ref{fig:initial_deform_simu}.
	Figure~\ref{fig:deloading_asym_simu}(a) is the results for deloding dynamics, where the detuning changes from $\delta/2\pi=-600$ Hz to $2$~kHz in 14.56 ms, followed by adiabatic turning off of Raman beams in 7 ms.
	$t_{\rm del}$ in Fig.~\ref{fig:deloading_asym_simu} is the time from when we start the deloading, and the deloding process ends at $t_{\rm del}=14.56 + 7=21.56$ ms. During the deloading process, a density dip appears in the projected order parameter at the phase winding point with winding $-2$ ($t_{\rm del}=4$ ms), which is then combined with one of the single vortices with winding 1, becoming a vortex with winding $-1$ ($t_{\rm del}=6$ ms). As a whole, a vortex-antivortex pair remains.
	During the residual time, the vortex configuration further changes and the distance between the vortex and antivortex becomes larger.
	When we deload to the final detuning of $600$~Hz and shorten the total period for the deloading process to 13.72 ms, the vortices in the final state come closer to each other and are located closer to the trap center [Fig.~\ref{fig:deloading_asym_simu} (b)].
	
	Figure~\ref{fig:initial_deform_exp} shows the experimental results
	of $\delta_{\rm del}/2\pi=2$ kHz (a) and a smaller final detuning 600 Hz (b) 
	corresponding to $m_F=-1$ component in the last row of Fig.~\ref{fig:deloading_asym_simu}~(a) and that in Fig.~\ref{fig:deloading_asym_simu}~(b), respectively. Being agreement with Fig.~\ref{fig:deloading_asym_simu}, the pair of density dips in Fig.~\ref{fig:initial_deform_exp} comes closer to the trap center and their distance becomes smaller for $600$~Hz than in the case of $\delta_{\rm del}/2\pi=2$~kHz.
	The experimentally measured $L_z\approx 0$ (Fig.~3 of the main text) indicates the two vortices are a vortex and an antivortex, also in agreement with the simulation. 
	Though Fig.~\ref{fig:deloading_asym_simu} (simulation) and Fig.~\ref{fig:initial_deform_exp} (experiment) are before and after time-of-flight, respectively, we have numerically confirmed that the vortex configuration is almost unchanged during the time-of-flight.
	
	\begin{figure}
		\includegraphics[width=7.0 in]{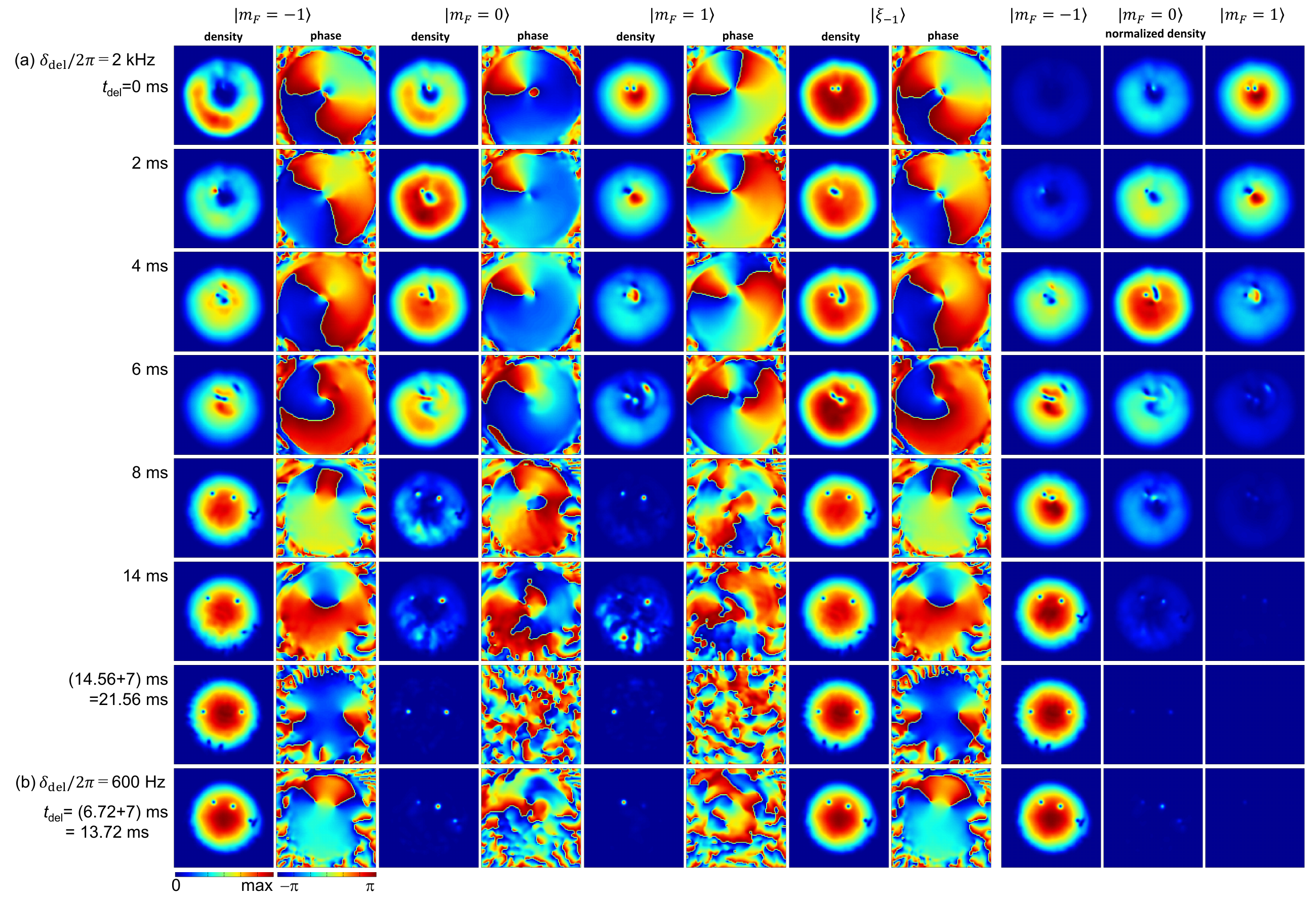}
		\caption{{\it in-situ} time evolution during deloding starting from $t_h=15$ ms in Fig.~\ref{fig:initial_deform_simu} calculated in 3D TDGPE simulation with asymmetry and energy dissipation. The meaning of each panel in the eight columns from the left is the same as that in Fig.~\ref{fig:initial_deform_simu}. The right-most three columns show the density profiles of $m_F=1,0,-1$ components in the same color scale at each time. During deloading, we sweep the detuning $\delta/2\pi=-600$ Hz to $\delta_{\rm del}/2\pi=2$ kHz in 14.56 ms (a) and to a smaller final detuning $600$ Hz in 6.72 ms (b) and then turn off the Raman beams in 7 ms. $t_{\rm del}$ is the time from when we start sweeping the detuning.}
		\label{fig:deloading_asym_simu}
	\end{figure}
	\begin{figure}
		\centering
		\includegraphics[width=3.0 in]{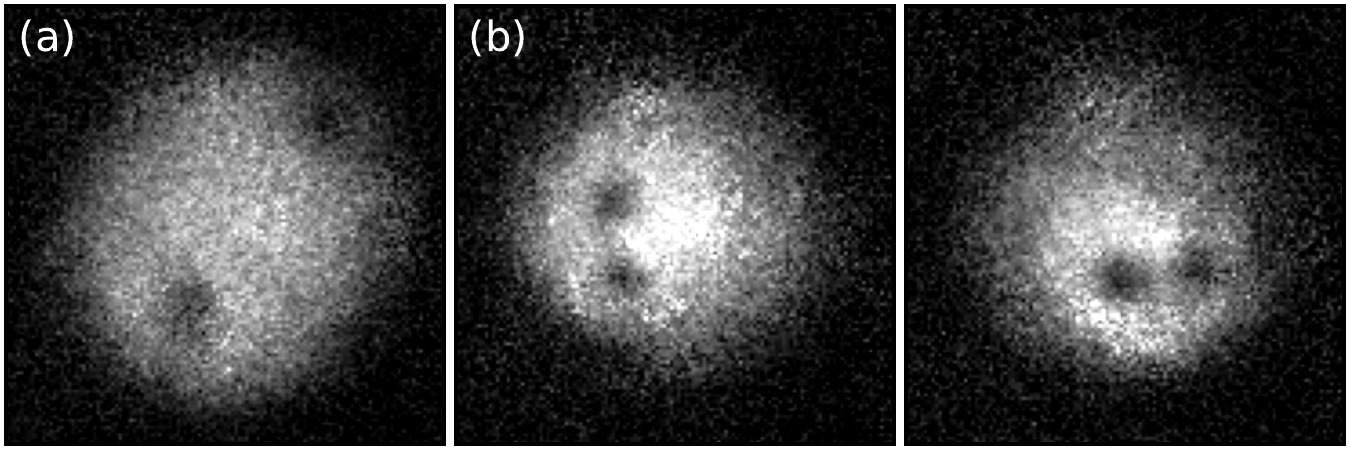}
		\caption{Time-of-flight images of the initial dressed state with $t_h \approx 15$~ms and $\delta/2\pi=-600$~Hz after deloading to $\ket{m_F=-1}$ with $\delta_{\rm del}/2\pi=2$~kHz (a) and $600$ Hz (b). The field of view is $168\micron\times168\micron$.}
		\label{fig:initial_deform_exp}
	\end{figure}
	
	At a longer $t_h$, we observe both experimentally and theoretically the cases when more than two vortices remain after deloading. Figure~\ref{fig:4vortex} shows an example of having four vortices, which is obtained by deloading to $\delta_{\rm del}/2\pi=2$~kHz starting from $t_h=45$ ms in Fig.~\ref{fig:initial_deform_simu}. 
	In the long-time dynamics, the system reduces $L_z$ by emitting some of the generated vortices and reaches the ground state with the aid of energy dissipation.
	\begin{figure}
		\centering
		\includegraphics[width=0.3\linewidth]{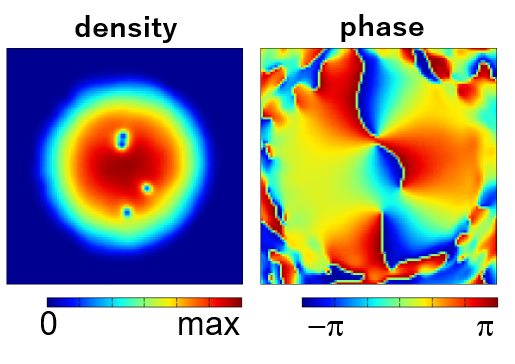}
		\caption{{\it in-situ} order parameter in $|m_F=-1\rangle$ in the $z=0$ plane after deloading to $\delta_{\rm del}/2\pi=2$ kHz starting from $t_h=45$ ms in Fig.~\ref{fig:initial_deform_simu}. This is the same data as Fig.~2h3 in the main text. In the numerical simulation, the asymmetry and energy dissipation of the system are included. Four vortices remain in the condensate, where two of them have phase winding $+1$ and the other two have $-1$.}
		\label{fig:4vortex}
	\end{figure}
	
	\begin{figure}
		\centering
		\includegraphics[width=6.75 in]{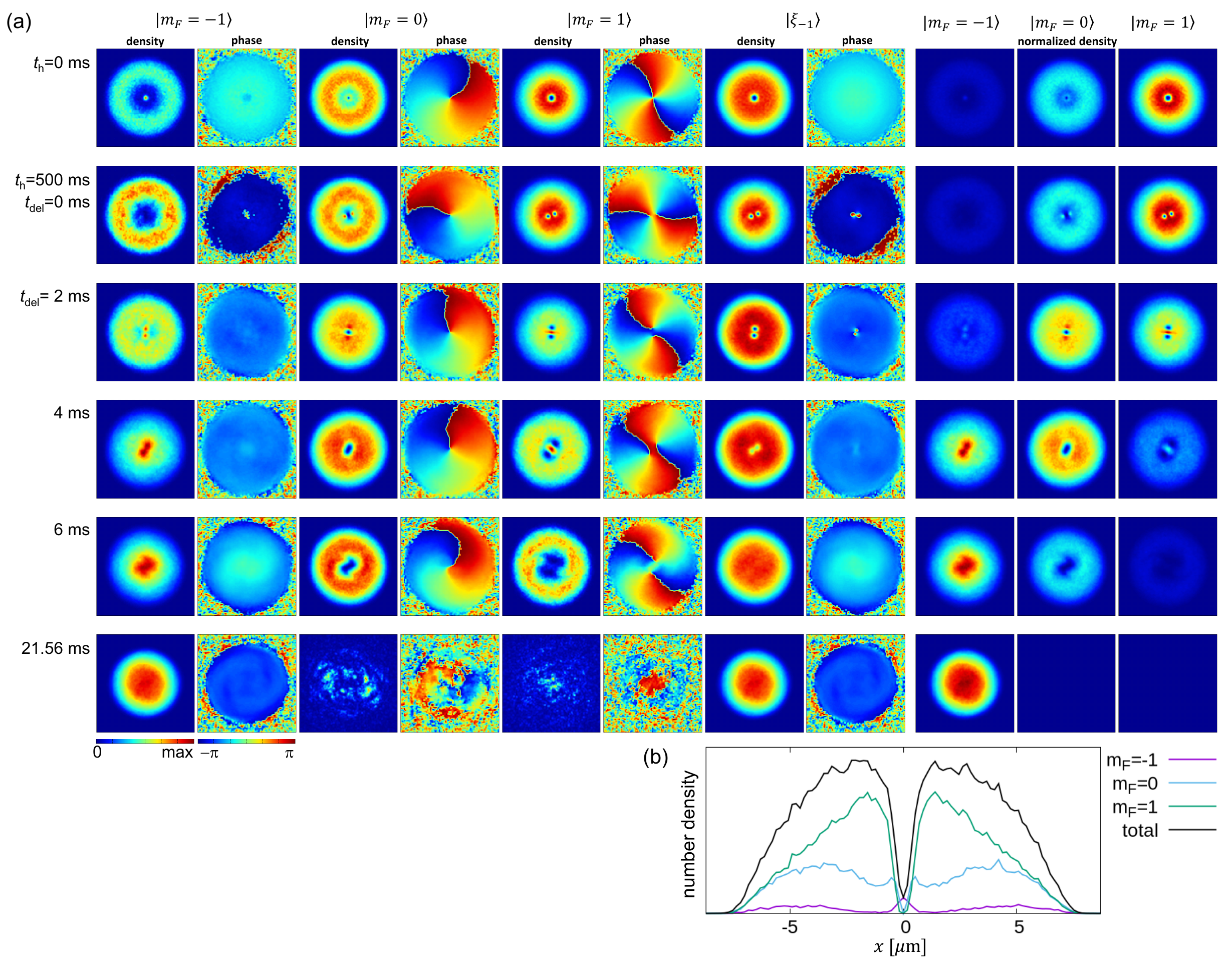}
		\caption{(a) Snapshots of 3D TDGPE simulation without asymmetry and energy dissipation. The meaning of each panel is the same as that in Fig.~\ref{fig:deloading_asym_simu}.	The top panels are the state just after loading the SOAMC ($t_h=0$ ms) to $\delta/2\pi=-600$ Hz, the next row shows the state at $t_h=500$ ms, at which we start the deloading process, and below are the time evolution during the deloading to $\delta_{\rm del}/2\pi=2$~kHz. (b) Density profiles of each bare spin component and the total density along the $x$ axis at $t_h=0$ ms.
		}	\label{fig:deloading_symmetric_simu}
	\end{figure}

	We note that in the above-explained dynamics, asymmetry as well as energy dissipation of the system are crucial.
	When the system is circularly symmetric, the vortex configuration is symmetric with respect to $r=0$. 
	In this case, it takes longer time for the splitting of a doubly quantized vortex in the $m_F=1$ component.
	Figure~\ref{fig:deloading_symmetric_simu} shows the vortex dynamics during the deloading process starting from $t_h=500$~ms in the absence of asymmetry and energy dissipation. In the panels of $t_{\rm del}=0$ ms, one can see that
	the initial deformation agrees well with that predicted by 2D BdG analysis (Fig.~\ref{fig:BdG_snapshots}), and 
	the vortex configuration is highly symmetric compared with, say, $t_h=45$~ms configuration in Fig.~\ref{fig:initial_deform_simu}, even though we hold the condensate in the dressed state much longer time. During the deloading process, the two density holes in the projected order parameter soon disappear by combining with the imprinted vortex with winding $-2$ at the LG beam center.
	Eventually, no vortex appears after deloading even at $t_h=500$ ms.

	\section{Azimuthal velocity profile}
	We show the azimuthal velocity for an atomic state whose spinor wave function is $\ket{\xi_{-1}}$ or the $\ell$-dependent $\ket{\xi_g}$ of the Gross-Pitaevskii (GP) ground state, respectively. Here we approximate our system as cylindrically symmetric.
	
	\begin{figure}
		\centering
		\includegraphics[width=3.5 in]{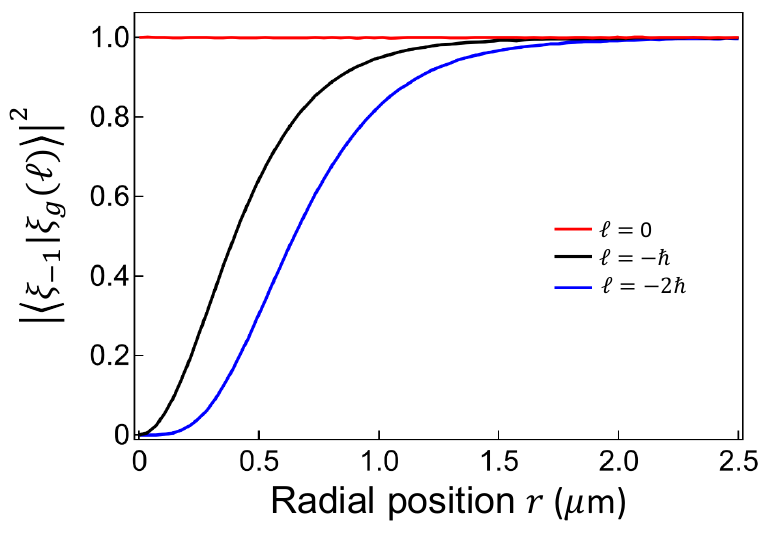}
		\caption{The square of the inner product $|\langle \xi_{-1} \ket{\xi_g(\ell=0,-\hbar,-2\hbar)}|$ for $\delta/2\pi=250$~Hz. Red, black and blue curves denote $\ell=0,-\hbar,-2\hbar$ for $\ket{\xi_g}$, respectively.}
		\label{fig:inner_product} 
	\end{figure}
	
	We decompose the three-component order parameter to a normalized spinor and an external wave function and suppose the external wave function is an eigenstate of the canonical OAM operator $\hat{\ell}$. The initial state has $\ell=0$ and is vortex-free. This state is adiabatically prepared in the lowest-energy dressed state and is almost identical to the GP ground state with $\ell=0$. On the other hand, the ground state for $\delta/2\pi<200$ Hz has $\ell=-\hbar$ or $-2\hbar$. We compute the spinor wave function $\ket{\xi_g(\ell)}$ for fixed $\ell=0,-\hbar,-2\hbar$, and compare them to $\ket{\xi_{-1}}$. Because of the kinetic-energy and quadratic Zeeman energy terms in the Hamiltonian, $|\xi_g(\ell)\rangle$ can deviate from $|\xi_{-1}\rangle$. For a not too small $\delta>0$, $\ket{\xi_{-1}}$ is close to $\ket{\xi_g}$ for $\ell=0$ provided $|\langle \xi_{-1} \ket{\xi_g}|^2 \approx 1$; see Fig.~\ref{fig:inner_product}. For $\ell=-\hbar,-2\hbar$, $|\langle \xi_{-1} \ket{\xi_g}|^2$ is small at small $r$, showing the deviation of $\ket{\xi_g,\ell=-\hbar,-2\hbar}$ from $\ket{\xi_{-1}}$. As for $\delta<0$, $\ket{\xi_{-1}}$ is close to $\ket{\xi_g}$ for $\ell=-2\hbar$. Due to a symmetry in the Hamiltonian, $|\langle \xi_{-1} \ket{\xi_{g}(\ell=0,\delta)}|=|\langle \xi_{-1} \ket{\xi_{g}(\ell=-2\hbar,-\delta)}|$.
	
	For a general state $\ket{\psi}=e^{i \ell\phi}\ket{\xi}$ in the gauge of Eq.~\eqref{eqn:xi_minus1}, the azimuthal velocity is
	\begin{align}\label{eqn:v_general}
		v(r)=\langle \psi|\frac{\hbar}{imr}\partial_{\phi}\otimes \textbf{1}|\psi \rangle,
	\end{align}
	which gives the same result of $v_{-1}$ in Eq.~\eqref{eqn:velocity} for $\ket{\xi}=\ket{\xi_{-1}}$.	The initial $\ell=0$ state and thus $v_{-1}(r)= -A_{-1}/m$. As $\delta$ decreases, $A_{-1}<0$ decreases monotonically and $v_{-1}(r)>0$ increases for any given $r$. This is just like the case of mechanically rotating BECs (before vortex nucleations) whose $\ell \sim 0$ (nonzero due to small asymmetry of the stirring potential) with $A= -m\Omega_{\rm stir}r$ and the velocity is $-A/m= \Omega_{\rm stir}r$ that increases with $\Omega_{\rm stir}$.
	
	We then use Eq.~\eqref{eqn:v_general} to compute the velocity of the GP ground state $v_g(r)$ for $\ell=0,-\hbar,-2\hbar$, respectively, and compare to $v_{-1}(r)$. For $\delta>0$, $-\hbar\leq rA_{-1}\leq 0$. Therefore, $v_{-1}(\ell=0)=-A_{-1}/m>0, v_{-1}(\ell=-\hbar)=-\hbar/mr-A_{-1}/m<0$ and $v_{-1}(\ell=-2\hbar)=-2\hbar/mr-A_{-1}/m<0$. Similarly, $v_{g}(\ell=0)>0$ and $v_{g}(\ell=-\hbar,-2\hbar)<0$. We compare $v_{-1}(r)$ and $v_g(r)$ at $\delta/2\pi=250$~Hz in Fig.~\ref{fig:velocity_vs_r}a, where it shows the absolute values of the velocities in the log scale.
	The unit of the dimensionless velocity is $\omega_r a_{\rm HO}=\sqrt{\hbar \omega_r/m}=0.00074$~m/s with $a_{\rm HO}=\sqrt{\hbar/m\omega_r}$. For $\delta>0$, $v_{-1}(r)$ largely agrees with $v_g(r)$ for $\ell=0$ even at small $r$. However, for $\ell=-\hbar,-2\hbar$, $v_{-1}(r)\propto 1/r$ near $r=0$, which deviates from $v_g(r)$, and this is consistent with the results in Fig.~\ref{fig:inner_product}. As $r$ decreases, $v_g(r,\ell=-\hbar,-2\hbar)$ stops increasing and decreases instead, showing a peak value at small $r=r_{\rm max}\sim 0.7\micron$. This results from the fact that $v_g(r=0)=0$ for the $\ell=0,-\hbar,-2\hbar$ dressed states, which are coreless vortex states, i.e., one of the bare spin $m_F$ component has zero OAM and contributes to nonzero density at $r=0$ with non-singular velocity $v=0$. 
	
	\begin{figure}
		\centering
		\includegraphics[width=6.8 in]{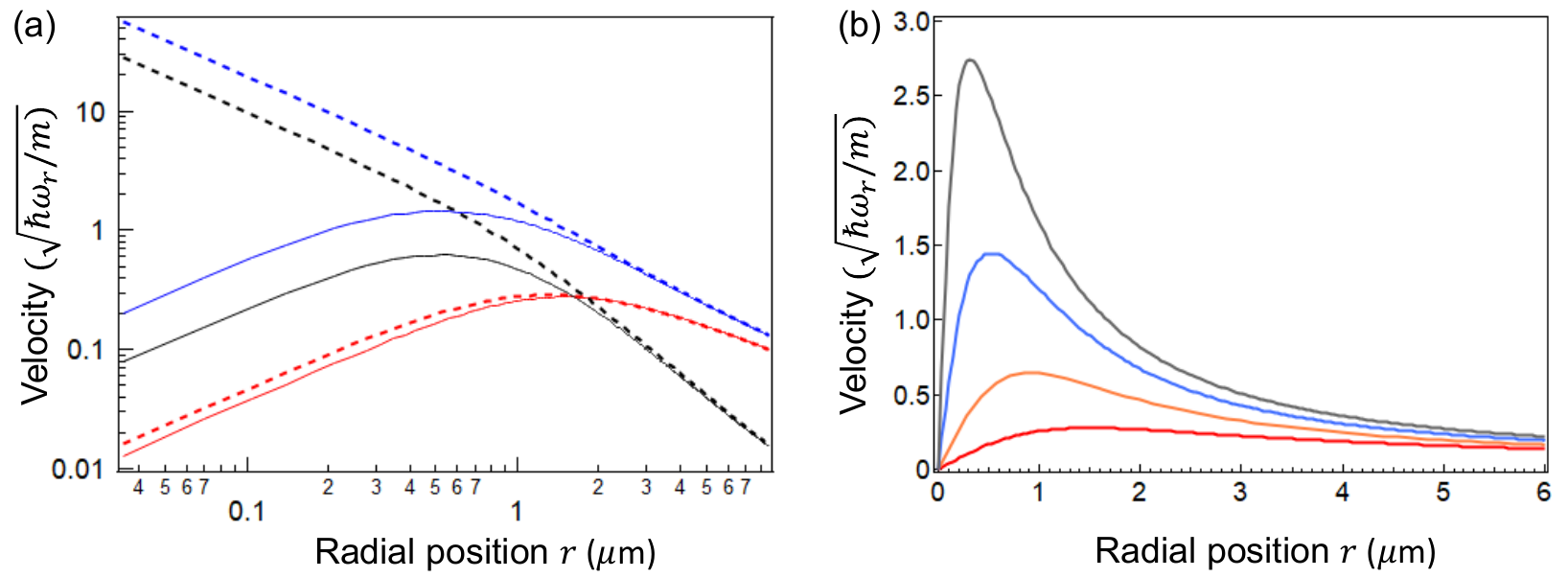}
		\caption{(a) Absolute values of the azimuthal velocity $v_{-1}(r)$ for $\ket{\xi_{-1}}$ (dashed) and $v_g(r)$ for $\ket{\xi_g}$ (solid) at $\delta/2\pi=250$~Hz for $\ell=0,-\hbar,-2\hbar$. The log scale plot shows $v_{-1},v_g$ for $\ell=0$ and $-v_{-1},-v_g$ for $\ell=-\hbar$ and $-2\hbar$. Red,black and blue curves denote $\ell=0,-\hbar,-2\hbar$, respectively. (b) Azimuthal velocity $v_g(r)$ of the Gross-Pitaevskii ground state with $\ell=0$ for detuning $\delta/2\pi=250$~Hz (red), $1$~Hz (orange), $-250$~Hz (blue),$-500$~Hz (grey).}
		\label{fig:velocity_vs_r} 
	\end{figure}
	
	Next we consider the detuning $\delta<0$. We plot $v_g(r,\ell=0)$ of our initial state prior to vortex nucleations for $\delta/2\pi=250,1,-250, -500$~Hz in Fig.~\ref{fig:velocity_vs_r}b. For all $\delta$, $v_g(r)$ has a peak at small $r_{\rm max}$ which decreases with $\delta$. According to the 3D TDGPE simulations, the dynamical instability appears when $\delta/2\pi \lesssim -200$~Hz, which indicates that negative energy excitations occur and the Landau criterion happens at $\delta/2\pi \gtrsim -200$~Hz. In Fig.~\ref{fig:velocity_vs_r}b the peak velocity of $\delta/2\pi= -250$~Hz is $\sim 1.5\omega_r a_{\rm HO}\sim 0.0011$~m/s.  We estimate the local sound velocity near the cloud center is about $0.0027$~m/s for our peak mean field energy $\sim 1.6$~kHz. We may argue that instability occurs when the peak of $v_g(r,\ell=0)$ equals to some numerical factor times the local sound speed.

	Here we discuss the calculations showing the existence of instability at detuning $\delta>0$, where one uses a high order LG beam to produce a Raman coupling with large $\Delta \ell/\hbar$. We choose the gauge of $\bar{\theta}+\bar{\gamma}=(\Delta \ell/\hbar)\phi$ such that the initial state prior to vortex nucleations is vortex-free with $\ell=0$ and the velocity $v_{-1}(r)=-A_{-1}(r)/m>0$ increases with decreasing $\delta$. Fig.~\ref{fig:vs_dl} shows $-rA_{-1}=\Delta \ell [1-\delta/(\Omega(r,\Delta \ell)^2+\delta^2)^{1/2}]$ and the velocity $v_{-1}$ vs. $r$ for various $\Delta \ell$ at $\delta/2\pi=200$~Hz. Here $\Omega(r,\Delta \ell)=\Omega_M e^{\Delta \ell/2} (r/r_M)^{\Delta \ell} e^{-\Delta \ell r^2/2r_M^2}$ is the Raman coupling strength for a general $\Delta \ell$. $v_{-1}$ has a peak value at the radial position $r_{\rm max}$ that is determined by $\delta$ and $\Delta \ell$. For $\Delta \ell=20\hbar$, the peak of $v_{-1}$ is $\sim 0.0009$~m/s and is comparable to that in our experiment when negative energy excitations occur. One can tune $r_{\rm max}$ to be small or large compared to the system size, which has unstable localized mode and surface mode, respectively. We then expect interesting competitions between the two physical mechanisms when $r_{\rm max}$ is comparable to the system size.
	
	\begin{figure}
		\centering
		\includegraphics[width=6.75 in]{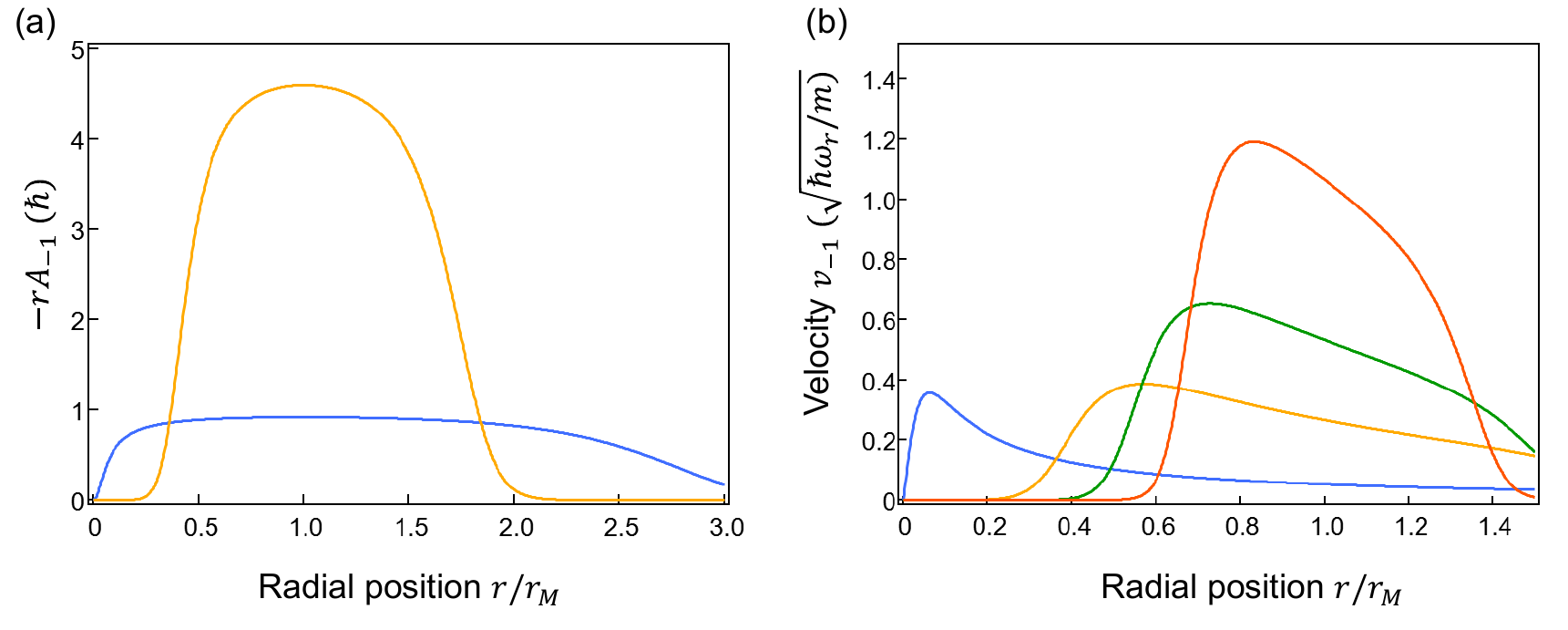}
		\caption{Simulations for the lowest energy dressed state $\ket{\xi_{-1}}$ with $\delta/2\pi=200$~Hz and the peak Raman coupling $\Omega_M/2\pi=2500$~Hz at $r=r_M$ for various $\Delta \ell$. (a) $-r A_{-1}$ vs. $r/r_M$ for $\Delta \ell=\hbar$ (blue) and $5\hbar$ (orange) under the gauge where the BEC is initially vortex-free. (b) Azimuthal velocity $v_{-1}$ vs. $r/r_M$ for $\Delta \ell=\hbar$ (blue), $5\hbar$ (orange), $10\hbar$ (green), and $20\hbar$ (red).}
		\label{fig:vs_dl} 
	\end{figure}

\end{document}